\documentclass[12pt,preprint]{aastex}
\usepackage{times,mathptm,graphicx} 

\slugcomment{\today } 
\shorttitle{ChaMP X-ray Point Source Number Counts} \shortauthors{Kim}
\begin{document}

\title{Chandra Multiwavelength Project \\
X-ray Point Source Number Counts and the Cosmic X-ray Background}

\author{Minsun Kim\altaffilmark{1,2},
Belinda J. Wilkes\altaffilmark{1},
Dong-Woo Kim\altaffilmark{1},
Paul J. Green\altaffilmark{1},
Wayne A. Barkhouse\altaffilmark{3},
Myung Gyoon Lee\altaffilmark{2},
John D. Silverman\altaffilmark{4},
and Harvey D. Tananbaum\altaffilmark{1}
}

\email{mkim@cfa.harvard.edu}
\altaffiltext{1}{Harvard-Smithsonian Center for Astrophysics,
60 Garden Street, Cambridge, MA 02138, USA}
\altaffiltext{2}{Department of Physics and Astronomy, Astronomy Program,
Seoul National University,
Seoul 151-742, Korea}
\altaffiltext{3}{Department of Astronomy, University of Illinois at Urbana-Champaign, 
Urbana, IL 61801, USA}
\altaffiltext{4}{Max-Planck-Institut f\"ur extraterrestrische Physik, D-84571 
Garching, Germany}

\begin{abstract}
We present the $Chandra$ Multiwavelength Project (ChaMP) 
X-ray point source number counts and the cosmic X-ray
background (CXRB) flux densities in multiple energy bands.
From the ChaMP X-ray point source catalog, $\sim 5,500$ 
sources are selected 
covering $9.6~deg^{2}$ in sky area. 
To quantitatively characterize the sensitivity and completeness of the ChaMP sample,
we perform extensive simulations. 
We also include the ChaMP+CDFs ($Chandra$ Deep Fields) number counts
to cover large flux ranges from $2\times10^{-17}$ to $2.4\times10^{-12}$
(0.5-2 keV) and from $2\times10^{-16}$ to $7.1\times10^{-12}$ (2-8 keV)
$erg~cm^{-2}~sec^{-1}$.
The ChaMP and the ChaMP+CDFs differential number counts are well fitted
with a broken power law.
The best fit faint and bright power indices 
are $1.49^{+0.02}_{-0.02}$ and
$2.36^{+0.05}_{-0.05}$ (0.5-2 keV), 
and $1.58^{+0.01}_{-0.01}$ and
$2.59^{+0.06}_{-0.05}$ (2-8 keV), respectively.
We detect breaks in the differential 
number counts and they 
appear at different fluxes in different energy bands.
Assuming a single power law model for a 
source spectrum, 
we find that the same population(s) of 
soft X-ray sources causes the break in the differential 
number counts for all energy bands. 
We measure the resolved CXRB flux densities 
from the ChaMP and the ChaMP+CDFs number counts with and
without bright target sources. 
Adding the known unresolved CXRB to the ChaMP+CDF resolved
CXRB, we also estimate total CXRB flux densities. 
The fractions of the resolved CXRB without target sources are
$78^{+1}_{-1}\%$ and
$81^{+2}_{-2}\%$
in the 0.5-2 keV and 2-8 keV bands, respectively,
somewhat lower, though generally consistent with earlier numbers
since their large errors.
These fractions increase by $\sim1\%$ when target sources
are included.
\end{abstract}


\section{Introduction}
What is the origin and nature of the cosmic X-ray background (hereafter CXRB)? 
Can detected X-ray sources account for the CXRB?
The CXRB consists of resolved and unresolved components. 
The resolved CXRB originates in 
discrete sources such as point and extended sources 
while diffuse components and 
faint sources which are below current flux limits  
contribute to the unresolved CXRB. 
The contribution of discrete X-ray sources to the CXRB can be directly 
measured from their number counts.
Using the deep surveys of $ROSAT$, $Chandra$, 
and $XMM$-$Newton$, 
the X-ray number counts have been determined down to 
flux limits of $\sim2.3\times 10^{-17}$ (0.5-2 keV), 
$\sim2.0\times 10^{-16}$ (2-8 keV), and
$\sim1.2\times 10^{-15}$ (5-10 keV) $~erg~cm^{-2}~sec^{-1}$
and $\thickapprox80-90\%$ of the CXRB is resolved
into discrete X-ray sources in the 0.5-2 keV and 2-8 keV 
(see \citet{bra05} for a detailed review).
In this study, using the $Chandra$ Multiwavelength 
Project (ChaMP)
and the $Chandra$ Deep Fields (CDFs) data, which include the largest 
number of sources and cover the widest sky area and flux range
from a single satellite $Chandra$ to date, 
we provide statistically robust X-ray number counts and   
CXRB flux densities without cross-calibration problem which 
is usually included in data from multiple satellites.
Also we study the X-ray number counts in multiple 
energy bands to systematically understand their behavior 
in each energy band. 
 
There have been many similar studies. 
Using the $Chandra$ survey of SSA13, \citet{mus00} presented 
the X-ray number counts in the 0.5-2 keV and 2-10 keV bands,
and suggested that detected hard X-ray sources account for
at least $75\%$ of the hard CXRB and that the mean X-ray spectrum 
of these sources is in good agreement with that of
the background.  
\citet{cow02} presented the 2-8 keV number counts from the 
CDF-S and CDF-N with SSA13/SSA22 and
\citet{ros02} presented those of the CDF-S,
finding that at most
$\sim 10\%$ ($\sim 15\%$) of the CXRB is unresolved in the soft (hard)
energy band.  
\citet{man03} presented 
the X-ray number counts in the 0.5-2, 2-8, and 0.5-8 keV
using the ELAIS survey data.
\citet{mor03} (hereafter M03) presented 
the X-ray number counts in the 0.5-2 and 2-10 keV
from combining data from three different surveys
($ROSAT$, $Chandra$, and $XMM$-$Newton$).  
They concluded that $95 \%$ and $89 \%$
of the soft and hard CXRB respectively can be resolved into discrete X-ray sources.
\citet{bau04} (hereafter B04) 
combined the CDF-N and CDF-S data
and measured the contributions of the faint X-ray source populations
to the CXRB. 
They found that  
$90\%$ (0.5-2 keV) and $93\%$ (2-8 keV) of the total CXRB 
was resolved into discrete sources.  
\citet{bas05} presented the number counts of the $XMM$-$Newton$/2dF
survey in the 0.5-2 and 0.5-8 keV bands and
\citet{chi05} presented the number counts of the $XMM$-LSS
survey in the 0.5-2 and 2-10 keV bands. 
\citet{wor05} found that the resolved fractions of the CXRB are $\sim85\%$ (0.5-2 keV),
$\sim80\%$ (2-10 keV), and $\sim50\%$ at $\gtrsim 8$ keV, respectively.
Recently, \citet{hic06} (hereafter HM06) directly 
measured the absolute unresolved 
CXRB from $Chandra$ Deep Fields images after excluding 
point and extended sources in those fields.
They also estimated the resolved X-ray source intensity from the CDFs and
from the number counts for brighter sources \citep{vik95,mor03}, and then
estimated the total CXRB flux density by combining the two.
They found that the resolved fractions of the CXRB are 
$77\pm3\%$ (1-2 keV) and $80\pm8\%$ (2-8 keV), respectively.  
Until now, using the $ROSAT$, $XMM$, and $Chandra$ data, 
these many studies have revealed that 
$\sim80\%$ of the CXRB is resolved into discrete X-ray sources in the
0.5-2 keV and 2-8 keV bands; however, the resolved fraction of the CXRB significantly 
decreases at $\gtrsim 8$ keV.   

The ChaMP is a serendipitous, wide area survey covering
intermediate and high fluxes using
$Chandra$ archival data.  
\citet{kim04a} presented the initial ChaMP catalog which contains $\sim800$ 
X-ray point sources in the central region of 62 of 149 ChaMP fields.
From the initial ChaMP catalog, \citet{kim04b} (hereafter KD04) presented 
X-ray number counts in the 0.5-2 keV and 2-8 keV bands.
To avoid the incompleteness of the selected fields, 
they selected sources having large X-ray source counts ($>20$) 
and located close to 
on-axis ($<400\arcsec$).  
The selected sample covered $\sim 1.1$ $deg^{2}$ in sky area.
In flux range from $10^{-15}$ to $10^{-13}$ (0.5-2 keV),
they detected the break in the differential number counts.
However, due to the shallow flux limit,
they could not detect the break in the 2-8 keV band.

In this study, we use the latest ChaMP X-ray point source catalog
which contains $\sim6,800$ X-ray point sources in 149 ChaMP fields 
with sky coverage area of $\sim 10 ~deg^{2}$ (Kim, M. et al. 2006,
hereafter KM06)
to determine the X-ray point source 
number counts in 6 energy bands.
To correct for incompleteness,
Eddington bias, and instrumental effects, and 
to include large off axis angles (up to $\sim15\arcmin$) and faint 
(down to $\sim 5$ source counts) sources,
we perform extensive simulations to calculate the sky coverage of
the selected sources as a function of flux. 
Using this large sample and the simulation results, 
we present the X-ray point source number counts
which fully cover the break flux in each energy band with 
small statistical errors.
Due to the wide flux range of the sample, 
we detect breaks in the differential number counts
in all energy bands and investigate what causes 
the different break flux in different energy band.
We also investigate the nature and the origin of the break in 
the differential number counts using hardness ratio 
HR ($=$(Hc$-$Sc)$/$(Hc$+$Sc), 
see Table \ref{tbl-energy-def} for energy band definition)
and redshift distribution of the X-ray sources. 
We also combine the ChaMP and CDFs (hereafter
ChaMP+CDFs) number counts to cover the full
available flux range.
From the ChaMP and the ChaMP+CDFs number counts,
we estimate the resolved CXRB flux densities in 6 energy bands.
By adding the known unresolved CXRB (HM06) 
to the resolved
ChaMP+CDFs CXRB flux density, we estimate the total CXRB flux densities 
in the 0.5-2 keV, 1-2 keV, and 2-8 keV bands.

In \S 2, we briefly describe the ChaMP data selection. 
In \S 3, we describe the method and results of the ChaMP simulations. 
In \S 4, the ChaMP and the ChaMP+CDFs number counts 
are presented in 6 energy bands
and are compared with previous studies.
In \S 5, we study the nature and origin of the break flux in 
the number counts. 
In \S 6, we estimate the resolved CXRB flux densities in 6 energy bands 
and the total CXRB flux densities in 3 energy bands.
In \S 7, the summary and conclusions of this study are shown.    
Throughout this study, quoted errors are for a $\pm1\sigma$
confidence level, unless otherwise noted.  
Although we perform this study in 6 energy bands 
(see Table \ref{tbl-energy-def}),
we only present the figures in the 0.5-8 keV, 0.5-2 keV, 
and 2-8 keV bands for simplicity; however, tables include the results 
in all energy bands. 
To compare with previous studies,
we assume photon indices of $\Gamma_{ph}=1.4$ 
and $\Gamma_{ph}=1.7$;
however, figures only with $\Gamma_{ph}=1.4$
are provided.  

\section{The ChaMP Sample Selection}
The X-ray point source sample is from 
the ChaMP X-ray point source catalog (KM06) 
which consists of $\sim 6,800$ X-ray sources 
in 149 $Chandra$ archival observations.
The ChaMP fields were selected to include ACIS observations at high
Galactic latitude, $|b|>20^{\circ}$.
Fields containing large extended sources, planetary objects, 
PI surveys, and local
group galaxies were excluded \citep{kim04a}. 
The ChaMP X-ray point source properties were obtained using 
a ChaMP-specific pipeline, XPIPE, which
uses $wavdetect$ 
\footnote{See http://cxc.harvard.edu/ciao.} detections
as source positions
and extracts source properties within a given aperture appropriate for
the local PSF size (a $95\%$ encircled energy radius at 1.5 keV) 
using $xapphot$ \citep{kim06b}. 

The ChaMP X-ray point source catalog is divided into main and supplementary
catalogs. 35 ChaMP fields overlap one another and the supplementary catalog 
contains sources from the 19 shorter exposure fields among these.
To avoid confusion due to duplicated fields,
our analysis uses the main ChaMP catalog which contains 130 ChaMP fields.
From the main ChaMP catalog, we selected sources in the I0, I1, I2 and I3 CCD chips
for 32 ACIS-I observations, and sources in I2, I3, S2, and S3 CCD chips
for 98 ACIS-S observations.
These sources are located within an off axis angle of $\sim15\arcmin$.
In addition, we selected sources with signal to noise ratio, $S/N>1.5$  
corresponding to source counts of $C\gtrsim 5$.
XPIPE detects sources in the B band
(0.3-8 keV, see Table \ref{tbl-energy-def} for energy band definitions)
and for all energy bands,
photometry is performed at the source positions determined in the
B band (see \S 3 in KM06).
Therefore, in our sample, it is possible to miss very soft (hard) sources
which might be detected only in the soft (hard) band
but not detected in the B band.
For sources with $S/N>1.5$
in the S (H) band,
the missing percentage of very soft (hard) sources is 
$5\%$ ($10\%$), when we assume matching of
all possible counterparts in the B and S (H) bands; however,
since we perform simulations to correct the incompleteness and bias in the 
ChaMP fields 
using the same detection technique as 
the ChaMP X-ray point source catalog (see \S 3), 
these very soft (hard) sources 
do not introduce an additional error in our number counts.

Since the ChaMP is a $Chandra$ archival survey, most ChaMP fields contain 
target sources selected by the PI and those targets are likely to 
be biased towards 
special X-ray populations
such as bright AGNs. 
Therefore, we excluded target sources to derive less-biased X-ray
number counts. Our selection results in $\sim 5,500$ sources in
the 0.3-8 keV band from
the ChaMP X-ray point source catalog. 
Table \ref{tbl-prop} lists 
the number of sources and the statistical
properties of the X-ray sources in each energy band.
Figure \ref{fig-count_flux} shows the counts and flux distributions of 
the final X-ray sample.
The median value of the distribution is also plotted.

\section{The ChaMP Simulations}
To determine accurate number counts, it is necessary to correct 
for the incompleteness of the sample as well as for instrumental
effects such as vignetting and the off-axis degradation of the PSF.
There are two major techniques to correct these biases, 
a semi-analytical approach,
and a Monte Carlo simulation. 
The semi-analytical approach is based on the flux limit map of a given field
which contains the faintest flux corresponding to the assumed significance level of
source detection \citep{joh03,cap05,chi05}.
This technique is efficient and reliable; however, it is possible
to undercorrect the incompleteness of the field because in this method 
the source detection probability is a function of only the source counts.   
The actual source detection probability in a $Chandra$ field is a complex function
of off axis angle and source counts: the detection probability
decreases as off axis angle increases and as source counts decrease (KM06). 
Therefore, to accurately determine the sky coverage of the ChaMP sample,
we performed extensive Monte Carlo simulations to correct  
incompleteness and biases of sample fields.

\subsection{Method}
The simulation method is described in detail in KM06 and
consists of three parts, (1) generating artificial X-ray sources with
MARX\footnote{See http://space.mit.edu/CXC/MARX/ and MARX 4.0 Technical Manual.},
(2) adding them to the observed image, and (3) detecting
these artificial sources with $wavdetect$ and extracting source
properties with the $xapphot$.
In step (2), we used the real $Chandra$ observations 
to accurately reflect the effects of background counts and source confusion 
in the ChaMP fields.

We performed simulations using all selected observations and 4 CCD chips 
in each observation (see \S 2).
We generated $1,000$ artificial X-ray sources per sample field
which corresponds to $\sim13,000$ artificial X-ray sources per $deg^{2}$. 
The number of sources in each field depends on
the effective exposure time of the observation and 
the neutral hydrogen column density, $N_{H}$,
toward the observed region of the sky.
On average, $11.7\%$ of the $127,178$ artificial
X-ray sources are detected in our simulations,
a total of $14,932$ artificial X-ray sources in 130 ChaMP fields.
The number of detected artificial sources is 2.5 times 
the $\sim6,000$ observed
sources and this number is statistically sufficient 
to estimate the properties of the ChaMP sample.

The form of the assumed number counts distribution
is not critical because we use the ratio
of input to output number of sources to determine
the sensitivity \citep{vik95,kim03}.
The actual X-ray differential number counts are described by a
broken/double power law
with a faint slope of $\sim-1.5$ and a bright slope of $\sim-2.5$
\citep{yan04,bas05,chi05}
in most energy bands; however, the break flux has not been well determined.
Therefore, we assumed a cumulative number counts distribution with a
single power law with a slope of $-1$ corresponding to 
a slope of $-2$ in the differential number counts,
taking the average of the faint and bright slopes
from the literature, in the 0.3-8 keV band.
From the assumed number count distribution, 
we selected the artificial source flux randomly. 
The artificial
source fluxes span from $5\times10^{-16}$
to $5\times10^{-10}~erg~cm^{-2}~sec^{-1}$ in the B band, covering the flux
range of the observed ChaMP X-ray point sources ($6\times10^{-16}$
$\sim$ $6\times10^{-12}$ $erg~cm^{-2}~sec^{-1}$).

The spectrum of the artificial sources was assumed to
be a power law ($F_{\nu}\propto \nu ^{-\Gamma_{ph}}$)  
with a photon index of $\Gamma_{ph}=1.7$, 
because the photon index of $\Gamma_{ph}$ for the observed ChaMP
sources spans $\Gamma_{ph}=1.5\sim2$ (KD04, KM06).
\citet{toz06} performed X-ray spectral analysis for 82 X-ray bright
sources in the CDF-S, and found a weighted mean value for
the slope of the power law spectrum is $<\Gamma_{ph}>\simeq1.75\pm0.02$.
The flux range of these bright sources in the CDF-S overlaps with
the faint flux end of the ChaMP sources, therefore,
we assumed that the faint ChaMP sources also have a photon index
of $\Gamma_{ph}\sim1.7$.
We assumed a Galactic absorption, $N_{H}$, \citep{sta92} for 
each observation; however, we did not include 
intrinsic absorption for the artificial source spectrum.
The spectrum of each X-ray point source was generated
using the XSPEC\footnote{See http://xspec.gsfc.nasa.gov/.} package.

The artificial source's position was randomly selected in each CCD chip area,
but was rejected if the source area at a given random position had
an exposure map value of less than $10\%$
of the maximum. 
This requirement is identical to that in the ChaMP X-ray point 
source catalog reduction procedure.
To avoid the over-crowding of the artificial sources,
$\sim250$ artificial sources per CCD were
divided into several groups to be added into the observed image:
while we did not allow the artificial X-ray point sources to overlap
one another,
we allowed overlap between artificial and real X-ray
sources to provide an estimate of source confusion in each observed field.
This resulted in $\sim10$ ($\sim20$) simulated images per ACIS-I (ACIS-S)
CCD, corresponding $\sim9,100$ CCD images (event files) to run $wavdetect$
($xapphot$).
Since $\sim11.7\%$ of the artificial sources ($\sim14,900$) are detected,
on average we added only $\sim1.6$ artificial sources
to each simulated image.
The net counts of the overlapping artificial sources with real sources were
corrected following the overlapping source correction methods described in
\S3.2.2 of KM06.

\subsection{Sky Coverage Area}
Using the results of the simulations described in \S 3.1, 
Figure \ref{fig-cflux} shows the number counts for artificial 
sources in the B band.  
The number count for sources, whose fluxes were randomly 
selected from the assumed
number counts ($magenta$ $solid$ $line$),
agree well with a slope of $-1$. 
However, there are slight statistical fluctuations 
at fluxes brighter than $10^{-13}$ $erg~cm^{-2}~sec^{-1}$.
due to small number statistics. The random sources
were selected per observation (see \S 3.1) and $1\sim2$ bright sources 
out of $1,000$ sources result in  
statistical fluctuations in each observation.
Also, since we fixed the flux maximum rather than using infinitely bright flux 
(see \S 3.1) for random sources,
the cumulative number of artificial sources drops at $\sim10^{-12}$ $erg~cm^{-2}~sec^{-1}$
rather than following a line of slope $-1$.
Since the aim of our simulations is to correct bias at faint fluxes, we do not 
require good statistics at bright fluxes. 
The number counts for artificial sources generated 
by MARX ($blue$ $dotted$ $line$) 
and that for
artificial sources detected by XPIPE ($red$ $dashed$ $line$) are also displayed.
The Eddington bias, that sources with
counts near the detection threshold will be preferentially detected
when they have upward fluctuations (e.g., \citet{ken03}), is evident
at faint fluxes ($S$ $<$ $10^{-14}$ $erg~cm^{-2}~sec^{-1}$)
in the simulated number counts.
Near $\sim 10^{-14}~erg~cm^{-2}~sec^{-1}$, the number of
detected artificial sources starts to decrease.

Figure \ref{fig-sc} displays sky coverage for sources with $S/N>1.5$ 
as a function of flux in 6 energy bands
assuming a photon index of 
$\Gamma_{ph}=1.4$.
The sky coverage area is the ratio of the number of detected over input
sources at a given flux, multiplied by the total sky area. 
The full sky area is $9.6~deg^{2}$. 
The geometrical area of a $Chandra$ CCD chip is $0.0196~deg^{2}$; however,
the net effective area is slightly larger due to the dither.
To accurately calculate the effective area,  
we follow the same method as $xapphot$:
all pixels in the exposure map were summed, 
excluding those pixels with
an exposure map value less than $10\%$ of the maximum
within the corresponding source area.
This criterion automatically excludes pixel positions
located near the edge of the CCD chip.

\section{X-ray Point Source Number Counts}
\subsection{The ChaMP Number Counts}
The cumulative number counts for sources brighter than a given flux $S$,
corrected by the corresponding sky coverage at $S$, is:
\begin{equation}
N(>S)=\sum_{S_{i}>S} {1 \over \Omega_{i}}, 
\end{equation}
where $S_{i}$ is the flux of the $i$th X-ray point source and $\Omega_{i}$
is the sky coverage that is the maximum solid angle covered 
by the flux $S_{i}$. 
Using the sources selected in \S 2 
and the corresponding sky coverage  
derived in \S 3.2, 
we derived the cumulative number counts for the ChaMP point sources. 
Since the differential number count is a derivative form of the cumulative number counts,
we derived the differential number counts from the cumulative number counts 
resulting from equation (1) as follows:
\begin{equation}
{\left.{dN \over dS}\right|}_i 
=-{{N_{i+1}-N_{i}} \over {S_{i+1}-S_{i}}}
\end{equation}
where $N_{i}$ is the cumulative source number at flux $S_{i}$.
Since the sky coverage rapidly decreases near the faint flux limit,
there are large statistical errors for the number counts 
in the faint flux regime.
Thus, for better statistics, we present the number counts brighter 
than the flux corresponding to
$10\%$ of the full sky coverage. 
For example, in the 0.5-8 keV band, this flux cut corresponds to 
$2\times10^{-15}$ $erg$ $cm^{-2}$ $sec^{-1}$ 
and $500$ sources fainter than this flux are not included in the final number counts.
In Figure \ref{fig-sherpa_fit_C},
we display the ChaMP differential number counts ($left~panels$) 
and cumulative number counts ($right panels$) in 3 energy bands.
Statistical errors on the number counts 
are assigned following \citet{geh86}. 

The shape of the cumulative number counts is curved rather than 
a single power law feature
and the differential number counts can be fit by a broken power law 
\citep{bal02,kim04b}
or by a double power law \citep{cow02,har03,yan04,chi05}.
Since errors for the cumulative number counts are not independent \citep{mur73},
it is difficult to estimate confidence levels of fitting parameters for the
cumulative number counts. 
Therefore, we fit the differential number count with a broken 
power law as follows:
\begin{equation}
{{dN} \over {dS}} = \left\{\begin{array}{ll}
K(S/S_{ref})^{-\gamma_{1}}, & S<S_{b}, \\
 & \\
K(S_{b}/S_{ref})^{(\gamma_{2}-\gamma_{1})}(S/S_{ref})^{-\gamma_{2}}, & S \ge S_{b}, \\
\end{array}
\right.
\end{equation}
where $K$ is a normalization constant and $S_{ref}$ is a normalization flux.
In this study, we set a normalization flux of $S_{ref}=10^{-15}$ 
$~erg~cm^{-2}~sec^{-1}$. 
$S_{b}$ is the break flux at which the slope of the differential 
number count changes. $\gamma_{1}$
and $\gamma_{2}$ are faint and bright power indices. 
The best fit parameters for the differential number counts 
are listed in Table \ref{tbl-fit_diff_nt} for 
photon indices of $\Gamma_{ph}=1.4$ and $\Gamma_{ph}=1.7$. 
The photon index $\Gamma_{ph}$ hardly affects 
$\gamma_{1}$ and $\gamma_{2}$, but it shifts $S_b$ somewhat. 
We display the best fit results
in the left panels of Figure \ref{fig-sherpa_fit_C}.
In all energy bands, we detected breaks and they 
appear at different fluxes in different energy bands. 
We discuss the break flux of the differential number count in more detail in \S 5.

By integrating equation (3), we can derive a formula for the cumulative number count
as follows:
\begin{equation}
{N(>S)} = 
\int{dN \over dS}~dS',
\end{equation}
therefore,
\begin{equation}
{N(>S)} = \left\{\begin{array}{ll}
K\left({1 \over {1-\gamma_{1}}} - {1 \over {1-\gamma_{2}}}\right) (S_{b}/S_{ref})^{(1-\gamma_{1})}
+K\left({1 \over \gamma_{1}-1}\right)(S/S_{ref})^{(1-\gamma_{1})},
& S < S_{b}, \\
K\left({1 \over \gamma_{2}-1}\right)(S_{b}/S_{ref})^{(\gamma_{2}-\gamma_{1})}(S/S_{ref})^{(1-\gamma_{2})}, & 
S \ge S_{b}.\\
\end{array}
\right.
\end{equation}
where definitions of parameters are the same as equation (3).
Using the best fit parameters derived from the differential number counts, we also
plot the best fit results for the cumulative number counts in the right panels of
Figure \ref{fig-sherpa_fit_C}. 

\subsection{The ChaMP+CDFs Number Counts}
To measure the discrete X-ray source contributions to the CXRB, 
it is important to derive the number counts over a wide range of flux. 
So far, M03 have presented the widest flux range of number counts
using a combination of three different surveys with
$ROSAT$, $Chandra$, and $XMM$-$Newton$.  
Due to the different calibrations of each satellite,
it is possible that additional systematic errors are introduced 
for this combined survey data. 
The ChaMP is a medium depth survey which covers the break flux regime 
in each energy band but with a faint flux limit too shallow to
estimate the resolved CXRB.
Therefore, to cover the faint flux regime as well, we decided to use the 
CDFs as well as the ChaMP data to determine 
the number counts.
Since the ChaMP and the CDFs are from the same satellite, $Chandra$, 
they provide number counts over a wide flux range without
systematic errors due to different calibrations.

The cumulative CDFs number counts (B04) are provided by
$Chandra~Deep~Field$ web site
\footnote{See http://www.astro.psu.edu/users/niel/hdf/hdf-chandra.html.},
and the corresponding sky coverage is from Figure 1 of B04.
Note that they combined the CDF-N and CDF-S source catalogs and then derived
the CDFs number counts. 
Using the cumulative CDFs number counts and their sky coverages, 
we derived the differential number counts
of the CDFs in the 0.5-2 keV and 2-8 keV bands. 
Then, we simultaneously fitted the differential
number counts of the ChaMP and the CDFs with a broken power law. 
In Figure \ref{fig-ncount_fit_wcdf}, we display the differential and the cumulative
number counts of the ChaMP+CDFs in the 0.5-2 and 2-8 keV bands.
The best fit parameters are listed in Table \ref{tbl-fit_diff_nt} 
and displayed in Figure \ref{fig-ncount_fit_wcdf} as red lines.
The ChaMP+CDFs number counts cover a flux range of $2\times10^{-17}$
$\sim$ $2.4\times10^{-12}$ (0.5-2 keV) and $2\times10^{-16}$
$\sim$ $7\times10^{-12}$ (2-8 keV) $erg~cm^{-2}~sec^{-1}$.
The bright flux end of the ChaMP+CDFs and the faint flux end of the ChaMP agree well. 
Thus the number counts are well established
with smaller statistical errors over a wide flux range.

Figure \ref{fig-sherpa_fit_6} compares the bet fits to the 
differential number counts
of the ChaMP alone with those of the ChaMP+CDFs. 
Overall, the ChaMP and the ChaMP+CDFs number counts agree within 
the uncertainties in the Sc and the Hc bands; 
however, in the Hc band, the faint power index $\gamma_1$ of 
the ChaMP ($1.83^{+0.16}_{-0.16}$)
is steeper 
than that of the ChaMP+CDFs ($1.59^{+0.13}_{-0.07}$)
at $1.2\sigma$ confidence.
B04 investigated the number counts of the CDF-N and the CDF-S independently,
as well as those of the combination of both CDFs, and found 
that in the Hc band the CDF-N is steeper than that of the CDF-S 
at flux fainter than $10^{-15}$ $erg~cm^{-2}~sec^{-1}$ and this deviation 
increases to $3.9\sigma$ at the faintest flux limits.
They suggested that this is caused by field-to-field variations, as also
reported by \citet{cow02}. Note that they did not find any significant
evidence for field-to-field variations
in the Hc band at fluxes brighter than $10^{-15}$ $erg~cm^{-2}~sec^{-1}$,
or across the entire flux range of the Sc band as already reported by KD04 in
the ChaMP study.   
Although the faint flux limit of the ChaMP number counts ($\sim2\times10^{-15}$
$erg~cm^{-2}~sec^{-1}$ in the Hc band) is brighter than 
that of the CDFs ($\sim2\times10^{-16}$), 
the large size of the ChaMP sample taken from
130 serendipitous $Chandra$ fields provides the best estimate of 
the average number counts. 
Therefore, it is likely that the CDF-S contains fewer faint sources
in the Hc band than the average number count distributions 
at $\gtrsim 1.2\sigma$
confidence levels. 

\subsection{Comparison with Previous Studies} 
In this section, we compare the ChaMP and the ChaMP+CDFs 
number counts with those of previous studies.
Table \ref{tbl-other} provides the best fit parameters,
the sky coverage, the faint and bright flux limits, 
the fitting space (in cumulative or differential spaces), 
and the fitting formula for each study.
Figure \ref{fig-area_other} shows the number of sources 
and the sky coverage of various surveys. 
The largest sky coverage area 
is 92 $deg^{2}$ and 74 $deg^{2}$ in the soft and hard band, 
respectively, for the combination data of $ROSAT$, $Chandra$, and
$XMM$-$Newton$ surveys (M03).
The ChaMP+CDFs covers the second largest sky area of 
9.8 $deg^{2}$; however, it contains the largest
number of sources due to the better resolution and 
sensitivity of $Chandra$ compared with
other X-ray observatories. 
In Figure \ref{fig-flux_other},
we plot the faint and bright flux limits of this and previous studies.
The ChaMP covers the widest flux range in the broad band.  
In the soft and hard bands, M03 covers 
the widest flux ranges, although the ChaMP+CDFs also spans a very 
wide flux range.
Overall, the ChaMP and the ChaMP+CDFs samples are 
second in sky area and flux range; however, 
they have the largest number of sources observed with a single satellite, 
$Chandra$.

The differential number counts can be described by 
a double power law \citep{cow02,har03,yan04,bas05,chi05} or
by a broken power law \citep{bal02,kim04b}.
\citet{man03} fitted their differential number counts 
with a single power law.
M03 introduced a fitting formula for the cumulative number counts 
which is a combination of two power laws (see equation (2) in M03),
and they fitted their
number counts in differential space as follows:  
\begin{equation}
{{dN} \over {dS}} = K(2\times10^{-15})^{\gamma_{1}} \left[{{\gamma_{1} S^{(\gamma_{1}-1)} 
+ \gamma_{2} S_{b}^{(\gamma_{1}-\gamma_{2})} S^{(\gamma_{2}-1)}} \over
{(S^{\gamma_{1}} + S_{b}^{(\gamma_{1}-\gamma_{2})} S^{\gamma_{2}})^2}}\right],
\end{equation}
where $\gamma_{1}$ and $\gamma_{1}$ are the two power indices,
$K$ is a normalization factor, and $S_{b}$ is
the discontinuity in the cumulative number counts space.
Therefore, we can not directly compare the exact 
parameters of equation (6) with those of a double power law 
or a broken power law. 

In Figure \ref{fig-gamma_other}, 
we compare the double or broken power law slopes 
of the differential number counts
for this study with those
for previous studies in the soft ($left$ $panels$) 
and the hard ($right$ $panels$) 
bands, respectively. 
In both soft and hard bands, the slopes at faint ($\gamma_1$)
and at bright ($\gamma_2$) fluxes 
for the ChaMP+CDFs agree with those of
previous studies within the uncertainties. 
We note that $\gamma_2$ of the ChaMP+CDFs is slightly steeper than that 
for the previous ChaMP study (KD04)
in which the hard band number counts was
fitted by single power law only for the bright flux regime
due to the shallow faint flux limit.
In this study, in the hard band, the $\gamma_1$ for the ChaMP is slightly steeper than
that for \citet{cow02}, H03, and the ChaMP+CDFs. 
Overall, the ChaMP+CDFs number counts 
agree with those of previous studies within the uncertainties in 
the soft and hard bands and they present statistically robust number counts
with the smallest uncertainties.

\section{Break of the Differential Number Counts}
\subsection{Origin of Different Break Fluxes in Different Bands}
In \S 4, we detected the break fluxes of the differential 
number counts in 6 energy
bands and they have different flux levels in each energy band 
(see $S_b$ in Table \ref{tbl-fit_diff_nt}).
The simplest explanation is that the break flux shifts as a function 
of energy band due to the corresponding
flux levels in each band. 
To investigate this possibility, we estimate the flux shift
by rescaling the break flux in a given energy band into 
the other energy bands using an assumed X-ray source spectrum.
We assumed a single power law model for the spectra
for the X-ray sources,
and estimated the expected break fluxes $S_{b,exp}$ in each energy band 
relative to a given break flux $S_{b,std}$ in a standard band
as follows:
\begin{equation}
S_{b,exp} (E_{1}-E_{2}) = S_{b,std} 
{{\int_{E_1}^{E_2}E^{-\Gamma_{ph}}EdE} \over 
{\int_{E_{S1}}^{E_{S2}}E^{-\Gamma_{ph}}EdE}},
\end{equation}
where $S_{b,std}$ is a break flux in a standard 
$E_{S1}-E_{S2}$ keV energy band and 
$\Gamma_{ph}$ is the photon index of a spectrum.
To calculate the expected break fluxes of the ChaMP and the ChaMP+CDFs, 
we used $S_{b,std}=$ $2.5\times10^{-14}$ and $S_{b,std}=$ $2.2\times10^{-14}$
$erg~cm^{-2}~sec^{-1}$ which are the measured break flux 
in the 0.3-8 keV band 
with a photon index of $\Gamma_{ph}=1.4$ and $\Gamma_{ph}=1.7$,
respectively. 

In Figure \ref{fig-bflux}, we compare the expected and measured break fluxes of 
the ChaMP, ChaMP+CDFs, and XMM-LSS \citep{chi05} number counts 
in several X-ray energy bands.
For the XMM-LSS, the expected break flux is calculated 
by assuming a photon index 
of $\Gamma_{ph}=1.7$ and $S_{b,std}$ is the measured break flux 
in the He (2-10 keV) band
for consistency with their study. 
Overall, expected and measured break fluxes agree within the uncertainties.
Since M03 fitted their differential 
number counts with a nonlinear equation
(see equation (6)) rather than a broken or a double power law,
we can not include their results. 
However, according to our own visual estimations,
the break fluxes of their study also follow the trend in Figure \ref{fig-bflux}.
Therefore, we conclude that 
the break flux shifts as a function of energy band due to the corresponding
flux levels in each band. Although we can not rule it out without detailed
source classification, which is beyond the scope of this paper, 
there is no need to invoke a
different population to explain the shift.

\subsection{Cause of the Break}
In \S 5.1, we found that different break flux levels in different 
energy bands could be explained by the identical X-ray population(s)
in each energy band.
Then, what causes the break flux? 
To answer this question, it is best to classify all X-ray sources
using optical spectroscopy and then to investigate which population(s)
is responsible for a break in their number counts.
However, it is difficult to obtain optical spectra 
of X-ray sources: some X-ray sources have very faint or no optical
counterparts. 
B04 classified their CDFs sources based on 
X-ray-to-optical flux ratio, optical spectrum, and X-ray 
properties such as X-ray spectrum and luminosity, and determined the
number counts for X-ray populations such as AGNs, star forming galaxies, 
and Galactic stars. They classified AGNs in more detail,
such as Type 1, Type 2, unobscured, and obscured AGNs, 
and they determined the number counts for each AGN subclass.  
However, the flux limits of the CDFs are not 
bright enough to investigate
the origin of break fluxes. 

The ChaMP is a multi-wavelength survey, including
follow-up at optical, spectroscopic, IR, and radio
wavelengths
as well as matching with published catalogs
such as SDSS and 2MASS.
Since these follow-up surveys are not yet completed,
only part of the ChaMP sample can be classified on the basis
of multi-wavelength properties.
The follow-up surveys of the ChaMP are still on going and we will be able to
investigate this issue in more detail by source classifications
covering break flux regimes.
Thus in this study we use only the X-ray properties such as the 
hardness ratio HR ($=$(Hc$-$Sc)$/$(Hc$+$Sc)) to
investigate the cause of the break flux and include 
all ChaMP X-ray sources.
The HR of the ChaMP sources was calculated by a Bayesian approach
which models the detected counts as a Poisson distribution
rather than a Gaussian distribution to successfully describe the  
statistical nature of the faint sources (Park et al. 2006; KM06).

\subsubsection{Hardness Ratio and Break Flux}
H03 constructed the differential number counts for the SEXSI 
sources in the 2-10 keV band divided into hard and soft sources
at a hardness ratio of HR$=0$.
They found that the number counts for the soft (HR$<0$) sources show a break
while the hard (HR$>0$) sources do not.
They suggested that, on average, the hard sources may be at lower redshift,
and so do not show the cosmological evolutionary effects which cause the
break. 
Following H03, we investigated the HR dependence of
the break flux for the ChaMP number counts in all energy bands.

The left panels of Figure \ref{fig-flux_hr} shows HR distributions 
of the ChaMP sources as a function of flux in 3 energy bands.
The break fluxes ($S_{b}$) reported in Table \ref{tbl-fit_diff_nt} of \S 4
are also plotted.
In the right panels of Figure \ref{fig-flux_hr}, we display
the number distributions 
of the HR for sources in the following categories:
all sources, sources fainter than the break flux ($S<S_{b}$), and
sources brighter than the break flux ($S>S_{b}$).
In all energy bands, there are fewer hard than soft sources 
at bright fluxes. 
We performed a Kolmogorov-Smirnov test (KS test, Press et al. 1992) 
to estimate the probability of faint and bright samples
having the same hardness ratio distribution 
and it is significantly low ($prob<1\times10^{-10}$)
in each energy band.
Overall, most hard sources are distributed at $S<S_{b}$,
while soft sources cover the entire flux range.   
Thus we defined samples at HR$\lessgtr0$ in all energy bands 
to investigate the relation between the source HR and 
the break in the number counts.
In \S 5.2.3, 
we investigate the flux-hardness ratio (S-HR) diagram in more
detail by performing a simple simulation for a test X-ray source
over a range of redshift and absorption  
to understand why bright, hard sources are rare in all energy bands. 

In the left panels of Figure \ref{fig-sub_hr_ran}, 
we display the differential number counts for
the soft (HR$<0$) and the hard (HR$>0$) sources in 3 energy bands.
We fitted the soft sources with a broken power law and
the hard sources with a single power law.
The best fit parameters are listed in Table \ref{tbl-fit_diff_nt_hr} and
displayed in Figure \ref{fig-sub_hr_ran} as red lines. 
In all energy bands, the differential number counts for soft  
sources show a break  
while those for hard sources do not. 
We performed a KS test \citep{pre92} for the flux distribution
of the soft and hard sources, and there is no possibility
that those samples have the same flux distribution 
($prob<4\times 10^{-17}$) in each energy band.
To statistically confirm the absence of the break in the hard 
source number counts, we also performed an F-test which is
a model comparison test to select the best model from two competing models,
a single and a broken power law. 
We used the $ftest$ in the $Sherpa$ 
\footnote{See http://asc.harvard.edu/sherpa/threads/index.html.} tool
and a standard criterion of $ftest$ for selecting 
the complex model is 
$significance < 0.05$ (the $95\%$ criterion of statistics).
We fitted the hard source number counts with both a single
and a broken power law and the broken power law model was   
rejected ($significance>6\times10^{-2}$) in all energy bands.
We note that for the soft source number counts, the single power
law model was rejected ($significance<5\times10^{-3}$) in all energy bands. 

We note that the number of soft sources is larger than that of
hard sources by a factor of $\sim10$ ($\sim2$) in the Sc (Hc) band
(see Table \ref{tbl-fit_diff_nt_hr}), thus it is possible that the
hard sources do not show the break due to small number statistics.
To check this possibility,  
we produced $1,000$ random subsets from the soft sources in each energy band: 
each subset has the same number of sources as the hard source samples. 
We derived the differential number counts for each subset
and display their averaged differential number counts 
in the right panels of Figure \ref{fig-sub_hr_ran}. 
The error bar represents the averaged error from each differential 
number counts.   
We note that the statistical fluctuation for each random subset 
is comparable to the averaged error.
Even with the reduced statistics, soft sources still show a detectable break.
Thus smaller number of hard sources does not prevent detection of a 
break in our sample. 
Our results agree with those reported by H03 in the 2-10 keV band.
Therefore, we conclude that the soft sources are 
responsible for the break in the differential number counts in
all energy bands.

We compare the best fit parameters of the soft and the hard 
subsamples with those 
of the total sample which includes all sources regardless of HR 
(see \S 4 and Table \ref{tbl-fit_diff_nt}). 
In Figure \ref{fig-comp_index} (a)-(c), 
we compare the soft sample with the total sample:
the faint power law indices are systematically 
shallower (at $5.7\sigma$) than those of the total samples, 
while the bright power law indices
and break fluxes agree well with those of the total samples, 
within the uncertainties. 
In Figure \ref{fig-comp_index} (d), we compare
the hard sample with the total sample:
the hard band (H and Hc) indices are 
shallower (at $2.6\sigma$) than those of the total samples, 
while the broad and soft band
indices agree to within the uncertainties. 
To quantitatively estimate the slope changes 
which indicate the strength of the break, 
we introduce a break factor as follows: 
\begin{equation}
BF\equiv(\gamma_2-\gamma_1)/(\gamma_2+\gamma_1),
\end{equation}
where $\gamma_1$ and $\gamma_2$ are the faint and bright power indices of
the differential number count (see equation (3)).
As the strength of the break increases, the break factor increases.
The break factors of the total and the soft sample  
are listed in Table \ref{tbl-break_factor}.
We found that  
the break factors tend to be smaller in the total samples than 
in soft samples for all energy bands,
and that the break factors tend to be larger  
in the soft bands than in the hard bands.  

\subsubsection{Redshift Distributions of Soft and Hard Sources}
Why do soft sources show a break while hard sources do not?
H03 suggested that the hard sources may be predominantly at
lower redshifts and so do not show the cosmological
evolution effects which cause the break.
To investigate this suggestion, 
we display the redshift distributions of the soft and hard sources 
in Figure \ref{fig-redshift_hr1}.
In our sample, 63 ChaMP fields  
were covered by optical follow-up survey
and 669 out of 5515 sources have redshifts,
of which we used the sources with $S/N>1.5$, matching with 
optical sources at the highest confidence level, and having
the highest confidence level of spectrum identification
(for detail descriptions of the optical follow-up survey, 
spectroscopy and redshifts in the ChaMP, 
see \citet{gre04} and \citet{sil05}).
In all energy bands, the hard sources distribute at lower  
redshifts than the soft sources.
We performed a KS test \citep{pre92}
to estimate the probability for soft and hard sources
having the same redshift distribution
and it is significantly low ($prob<1\times10^{-4}$)
in each energy band.

Since the spectroscopy of the ChaMP sources was biased 
toward optically bright sources, this bias may affect the X-ray source 
selection for measuring redshifts and may cause the lower redshift
distribution of hard sources: 
more soft than hard sources selectively have
redshifts. 
In the top panel of Figure \ref{fig-imat_hr}, we display the hardness ratio 
distributions of the ChaMP sources
in the B (0.3-8 keV) band in the following categories:
all sources,
sources with optical imaging observations in 63 ChaMP fields,
sources having an optical counterpart,
and sources having a redshift.
In the bottom panel of Figure \ref{fig-imat_hr},
we display the number ratios of the last three subsamples
over total sample in each hardness
ratio bin. Overall, $60\%$ of sources were covered by the optical
follow-up survey, of those, $32\%$ of sources have an optical counter part,
and of those, $5\%$ of sources have a redshift. 
The fraction of sources having a redshift is 
$5.2\pm2\%$ for soft sources
and $6.7\pm6\%$ for hard sources, respectively.
Thus, the bright source selection in the optical band 
does not significantly affect the 
X-ray source selection for measuring redshift 
as a function of hardness ratio.

Then, why are hard sources distributed at lower redshifts? 
Since the QE and the effective area
of the $Chandra$ ACIS are lower and smaller in the hard band
\footnote{See Chapter 6 of the $Chandra$ Proposers' Observatory Guide Rev.8.0.},
and the X-ray source counts are fewer in the hard band than in the soft band
when a power law spectrum is assumed,
it is possible to miss more hard sources than soft sources especially
at higher redshift. Also, it is possible that an intrinsically hard source
may be observed as a soft source due to the cosmological redshift.
In \S 5.2.3, we quantitatively investigate this issue in more detail.

\subsubsection{Redshift and Absorption Effects on X-ray Properties}
To understand the dependence of X-ray properties, 
such as flux and hardness ratio, on the redshift 
and absorption,
we performed a simple simulation for a test X-ray source
using the $Sherpa$ 
\footnote{See http://asc.harvard.edu/sherpa/threads/index.html.}
tool.  
We assumed a power law model spectrum for the test X-ray source as
follows:
\begin{equation}
F(E) = K (E\times(1+z)/(1 keV))^{-\Gamma_{ph}},
\end{equation}
where $z$ is a redshift and $\Gamma_{ph}$ is a photon index of the
test X-ray source. $K$ is a normalization constant at $1$ keV 
in units of $photons$ $keV^{-1}$ $cm^{-2}$ $sec^{-1}$ and we set $K=0.5$.  
We assumed a Galactic absorption and intrinsic absorption 
using Wisconsin cross-sections \citep{mor83} as follows:
\begin{equation}
A(E) = exp(-N_{H,Gal}\times \sigma(E)),
\end{equation}
\begin{equation}
A(E) = exp(-N_{H,int}\times \sigma(E\times (1+z))),
\end{equation}
where $\sigma(E)$ is the photo-electric cross-section not 
including Thomson scattering and $z$ is the redshift.
$N_{H,Gal}$ and $N_{H,int}$ are
equivalent hydrogen column density in units of $atoms$ $cm^{-2}$
for the Galactic and intrinsic absorption, respectively.
We selected a ChaMP ACIS-I observation whose Galactic absorption is
$N_{H,Gal}=1.18\times 10^{20}$ $cm^{-2}$. Using the 
ancillary response function (ARF) and 
redistribution matrix function (RMF) files,  
we calculated the source flux and hardness ratio at the aim point
for various ranges of redshift ($0\leq z \leq 10$) and intrinsic absorption
($10^{20} \leq N_{H,int} \leq 10^{24}$ $cm^{-2}$). 

In Figure \ref{fig-hr_flux}, we display the 
flux-hardness ratio (S-HR) diagram in 3 energy bands.
All ChaMP sources with $S/N>0$ are displayed and 
the grid indicates the predicted
location of a test source 
with various redshifts ($z=$0, 1, 2, and 3)
and intrinsic absorption column densities
($logN_{H,int}=$20, 21.7, 22, 22.7, and 23.7). 
A photon index of $\Gamma_{ph}=1.4$ was assumed for the test source
spectrum.
We note that the flux of grid was renormalized to be displayed with the ChaMP
scatter plot.
The source becomes fainter with increasing absorption and 
with increasing redshift.
The source becomes harder with increasing intrinsic absorption,
but softer with increasing redshift.
In the soft band (0.5-2 keV), this effect is more significant
than in the hard and broad bands.
From this ideal case study, 
we can understand the observed flux-hardness ratio diagram 
in which there are fewer bright hard sources in each energy band
(see \S 5.2.1).
The test X-ray source does not
cover the region HR$\leq -0.4$; but will cover this regime when 
a steeper power law index (i.e., $\Gamma_{ph}>2$) is assumed
(see Figure \ref{fig-hr_z}).  

Figure \ref{fig-hr_z} shows the hardness ratio of
the test X-ray source as a function of redshift with a range of
intrinsic absorption ($20\leq logN_{H,int} \leq 24$) for different
photon indices.
The test source with steeper power index 
(i.e., $\Gamma_{ph}=2$ and 3)
covers the soft hardness regime (HR$\leq-0.5$).
Again, the test source becomes harder with increasing intrinsic absorption; 
but, softer with increasing redshift.
For example, in the top left panel (assumed $\Gamma_{ph}=1.4$), 
the hard source with $logN_{H}=22$ is not observed
as a hard source anymore even at $z\sim1$,
and most hard sources are observed as soft sources at $z>3$. 
Therefore, a hard source with high redshift
is observed as a soft source in the observed frame due to the 
cosmological redshift,  
and so hard sources with high redshifts are rare (\S 5.2.2).
Thus the hard source number counts do
not include high redshift hard sources, 
while the soft source number counts include
both intrinsically hard and soft sources. 

In \S 5.2, we found that the hard sources do not show a break
in their number count distributions and distribute at lower
redshifts compared to soft sources.
The soft sources show the break in their number count distributions,
and distribute from low to high redshifts (see Figure 12 and 14).
The observed soft sources may be a mixture of soft sources and redshifted
hard sources (see Figure 17).
These results likely support the suggestion that the hard sources
may be preferentially at lower redshifts, and so do not show 
cosmological evolution effects (H03).
In addition to H03's suggestion,
we suggest that the break in the soft source number counts may be caused
by the mixture of X-ray source populations as well as cosmological
evolution effects.
To investigate this suggestion,
we need redshifts/classifications of the X-ray sources.
Since it is not possible to speculate on the distribution of
properties such as intrinsic absorption $N_{H,int}$ from the source counts alone,
we need to assume a model for the $N_{H,int}$ distribution of X-ray point sources
as a function of redshift and luminosity or perform X-ray spectral analysis.
B04 found that the source density of Type 1 AGNs
is 10-20 times lower than that of Type 2 AGNs at the CDF flux
limits in both of the 0.5-2 keV and 2-8 keV bands.
Also, they found that the source density of unobscured/mildly
obscured AGNs is 2-3 times lower than those of obscured AGNs
at the CDF limits.
\citet{laf05} found that the fraction of absorbed ($N_{H}>10^{22}
cm^{-2}$) AGNs decreases with intrinsic X-ray luminosity but increases
with redshift.
The fraction of Type 1/Type 2 AGN (absorbed/unabsorbed AGN)
probably affects the break in the differential number counts.
Also, since the hard band is less affected by absorption than the
soft band, it is possible that
the strength of the break is related to the fraction of absorbed sources
in each energy band.
We expect that further studies can be performed using the ChaMP
data once we have more optical/spectroscopy follow-up observations.

\section{Cosmic X-ray Background} 
\subsection{Resolved Cosmic X-ray Background Flux Density} 
The contribution of discrete sources to the CXRB flux density 
can be calculated from 
the differential number counts as follows (M03):
\begin{equation}
F_{resol}=\int_{S_{faint}}^{S_{bright}} \left({dN} \over {dS}\right)S' dS',
\end{equation}
where $S_{faint}$ and $S_{bright}$ are the faint and bright flux limits
of the sample. 

The ChaMP is a serendipitous $Chandra$ archival survey, therefore most observations
contain target sources as intended by the PI and 
which have brighter flux than non-target sources 
as shown in Figure \ref{fig-flux_target}.      
To avoid biased source selection,
we excluded target sources for deriving the ChaMP and the ChaMP+CDFs
number counts.  
Even though we have only 85 target sources in total, their contributions to 
the CXRB flux density are not negligible because of their brightness (M03).
Thus we need to correct the bright target source contributions  
to the CXRB. Since the target sources cover a relatively
wide flux range, $3\times10^{-16}\sim 7\times10^{-13}$ (0.5-2 keV) and
$2\times10^{-14}\sim 2\times10^{-12}$ (2-8 keV) $erg$ $cm^{-2}$ $sec^{-1}$, 
respectively, we can not simply adapt the bright part of the number
counts for full sky surveys such as $ROSAT$ $All$ $Sky$ $Survey$ (soft band) or
$HEAO$-1 A2 extra galactic survey (hard band) that were used by M03
to correct their bright target source contributions to the CXRB. 
Therefore, we present the lower and upper limits of the resolved CXRB flux density
from the ChaMP and ChaMP+CDFs number counts by excluding 
target sources and including target sources, respectively.
We derived again the ChaMP and the ChaMP+CDFs number counts including
target sources and list their best fit parameters 
in Table \ref{tbl-fit_diff_yt}. Comparing with the best fit parameters without
target sources (Table \ref{tbl-fit_diff_nt}), 
target sources make the bright power law indices
($\gamma_{2}$) shallower at $0.9\sigma$ confidence, while the faint power
law indices ($\gamma_{1}$) and break fluxes ($S_{b}$)
show differences at $0.5\sigma$ and  $0.3\sigma$, respectively.  
 
Table \ref{tbl-cxrb_resol} lists the resolved CXRB flux densities 
and their contributions to the total CXRB 
from the ChaMP and the ChaMP+CDFs when the target sources are excluded 
or included in 6 energy bands. 
The average total CXRB flux densities of
$(7.52\pm0.35)\times 10^{-12}$ (0.5-2 keV)
and of $(1.79\pm0.11)\times 10^{-11}$ (2-8 keV) $erg$ $cm^{-2}$ $sec^{-1}$ $deg^{-2}$
(B04) are assumed.
In the Bc band, the total CXRB flux density is the sum of 
those in the Sc and Hc bands.
The total CXRB flux density in the B, S, and H bands are rescaled from 
the Bc, Sc, and Hc bands, respectively,
by assuming a photon index of $\Gamma_{ph}=1.4$.
In Figure \ref{fig-cxrb_SC_HC},
we display the resolved CXRB flux density calculated from the ChaMP+CDFs as a
function of flux limit in the Sc and Hc bands, respectively.
We plot the resolved CXRB flux densities calculated from the
differential number counts
with ($blue$ $lines$) and without ($red$ $lines$) target sources.
The ChaMP sources resolve the total CXRB without (with) target sources
at $80\pm2(86\pm2\%$) and $72\pm2(76\pm2\%$) 
in the Sc and Hc bands, respectively.
Since the ChaMP+CDFs covers wider flux range than the ChaMP,
the ChaMP+CDFs sources resolve more total CXRB by up to
$4\sim7\%$ in each band. 
We extrapolated the best fit ChaMP+CDFs number counts without target sources
down to $10^{-20}$ $erg$
$cm^{-2}$ $sec^{-1}$ and found that the total CXRB is not fully
resolved in the soft and hard energy bands within the uncertainties.
We note that for the 2-8 keV band, extrapolating the best fit
ChaMP+CDFs number counts with target sources down to
$\sim10^{-17}$ $erg$ $cm^{-2}$ $sec^{-1}$,
the total CXRB flux density is fully resolved within the large uncertainties.
In Figure \ref{fig-cxrb_6_comp}, we display the difference between the resolved
CXRB excluding and including target sources, normalizing with that of excluding targets.
At bright flux limits,
the flux density differences are
up to $100\%$; however, at faint flux limits,
the differences are less than $10\%$
in each energy band.
The resolved CXRB with and 
without target sources are upper and lower limits of the resolved CXRB, respectively,
and the actual resolved CXRB is between those values.  
In all energy bands, the fractions of the resolved CXRB increase by $5\sim6\%$ 
when target sources are included.

In Figure \ref{fig-fmin_cxrb}, we display the resolved CXRB 
flux density ($top$) and
the fraction of resolved CXRB ($bottom$) 
as a function of faint flux limit in each
energy band. 
Excluding bright target sources, 
M03 estimated the resolved CXRB flux densities to be
$0.69^{+0.03}_{-0.02} \times 10^{-11}$ (0.5-2 keV) and 
$1.40^{+0.09}_{-0.08}\times 10^{-11}$ (2-8 keV,
rescaled from that in the 2-10 keV band
assuming a photon index of $\Gamma_{ph}=1.4$) 
$erg$ $cm^{-2}$ $sec^{-1}$ $deg^{-2}$, respectively. 
From the ChaMP+CDFs without target sources, 
we estimated the resolved CXRB flux density
to be $0.63^{+0.01}_{-0.01} \times 10^{-11}$ (0.5-2 keV), lower than
that of M03 at $2\sigma$ level, 
and $1.40^{+0.03}_{-0.03}\times 10^{-11}$ (2-8 keV)
$erg~cm^{-2}~sec^{-1}~deg^{-2}$ in good agreement with M03, respectively.
The fractional contributions of 
the ChaMP+CDFs X-ray point sources excluding (including) 
target sources to the total CXRB are 
$84\pm2(91\pm2)\%$ and
$78\pm2(84\pm4)\%$
in the Sc and Hc band, respectively.

\subsection{Total Cosmic X-ray Background Flux Density} 
In \S 6.1 we used the measured total CXRB flux density (B04);
however, we can also derive the total CXRB flux density 
from the sum of resolved and unresolved components, using the 
resolved CXRB estimated from the ChaMP+CDFs.
Recently, HM06  
measured unresolved CXRB flux densities using the $Chandra$ Deep
Fields North and South of
$(0.18\pm0.03)\times 10^{-11}$,  
$(0.34\pm0.17)\times 10^{-11}$, and 
$(0.10\pm0.01)\times 10^{-11}~erg~cm^{-2}~sec^{-1}~deg^{-2}$ 
in the 0.5-2 keV, 2-8 keV, and 1-2 keV bands, respectively, 
after removing all detected point and extended sources in those fields.
They also estimated the resolved CXRB flux densities from the CDFs, 
and from the $ROSAT$ (0.5-2 keV, \citet{vik95}) and
the $Chandra$, XMM, and ASCA (2-10 keV, M03)
for the flux ranges brighter than CDFs. 
And then, they derived the total CXRB flux densities by adding those 
resolved and unresolved components.

We derive the total CXRB by adding HM06's
unresolved CXRB values to the resolved CXRB
of the ChaMP+CDFs.
Since these are estimated from a single satellite, $Chandra$,
there are no cross-calibration uncertainties as in multiple satellite data.
In Table \ref{tbl-cxrb_total}, we list the resolved, unresolved, and total 
CXRB flux densities estimated from this and previous studies. 
For this study, the resolved CXRB in the 1-2 keV band was rescaled
from that in the 0.5-2 keV assuming a photon index of $\Gamma_{ph}=1.4$.
We also provide the total CXRB with and
without target sources, which gives 
lower and upper limits to the total CXRB, respectively.
The actual total CXRB is between these two values.
The total CXRB flux densities increase by $\sim6\%$ when target
sources are included.
Our results agree well with those of HM06 but are lower than earlier numbers,
$\sim80\%$ compared with $90-94\%$ (M03; B04, see Table \ref{tbl-cxrb_total}).
Given the large uncertainties in the M03 and B04 studies, they remain
marginally consistent ($\sim2\sigma$ differences). 

We note that, in this study, the total CXRBs include two kinds of unquoted 
uncertainty.
First, the total CXRBs could be overestimated due to the incompleteness correction.  
The number counts are corrected for incompleteness;
however, this corrected portion may be also included in the unresolved CXRB 
since they are not resolved in the observations.
Our resolved CXRBs were corrected for incompleteness by $7\%$ (0.5-2 keV) $\sim$ 
$18\%$ (2-8 keV);
however, since HM06 used only the
central $5\arcmin$ around each CDF pointing in which the count recovery
rate and the detection probability of the source are higher 
than those at the off-axis region (B04; KM06),
the duplicated fraction of the total CXRB is much smaller than the corrected fraction.       
Second, the total CXRBs could be underestimated since 
we do not include the resolved CXRB
that originates from X-ray extended sources.
The resolved CXRB from the $ROSAT$ deep cluster survey
in the flux range of $10^{-14}$$\sim$$10^{-11}$ $erg$ $cm^{-2}$ $sec^{-1}$
\citep{ros98}
increases our total CXRB flux density by up to $10\%$,
and their contribution to the total CXRB 
will be $9.5\%$ in the 0.5-2 keV band. 
We note that, with this extended source contribution, 
our total CXRB still agrees with that of other studies within the uncertainties.
Meanwhile, so far, there are no number counts for X-ray extended sources 
in the hard band.
Since the spectrum of an extended source is not a simple power
law, we can not rescale the resolved CXRB in the soft band to 
that in the hard band. Thus, we did not include the extended source contribution
to the total CXRB in all energy bands. 
Since the ChaMP includes extended sources as well \citep{bar06},
in a future ChaMP study 
we expect to determine their number counts both in the soft 
and the hard bands with higher confidence levels 
by performing extensive simulations to accurately correct their incompleteness.
Then, we will be able to estimate the resolved CXRB for extended sources,  
giving us a self-consistent total CXRB flux density from $Chandra$.       

\section{Summary and Conclusions}
We present the $Chandra$ Multiwavelength Project (ChaMP)
X-ray point source number counts
in 6 energy bands.
We also present the ChaMP+CDFs number counts in the 0.5-2 keV and
2-8 keV which covers large flux ranges with small statistical errors. 
Using these number counts, we measure the resolved and total 
CXRB flux densities in multiple X-ray energy bands.
The main results and conclusions of this study are the following. 

1. The number counts of the ChaMP and the ChaMP+CDFs are well fitted 
with a broken power law.
The best fit faint and bright power indices of the ChaMP+CDFs 
are $1.49^{+0.02}_{-0.02}$ and
$2.36^{+0.05}_{-0.05}$ (0.5-2 keV), 
$1.58^{+0.01}_{-0.01}$ and
$2.59^{+0.06}_{-0.05}$ (2-8 keV), respectively.
The number counts in this study agree with 
those of previous studies within the uncertainties but are better
constrained.

2. In all energy bands, we detect a break in the differential 
number counts which is a function of 
energy band.  
The origin of the break depending on energy band 
can be explained by the identical X-ray population(s) 
in each energy band.

3. In all energy bands, the soft sources are responsible for 
the break in the differential number counts. 
A hard X-ray source becomes softer with increasing
redshift, and so the hard source number counts do not 
include high redshift sources while the soft source number counts
include both soft sources with full range of redshifts
and intrinsically hard sources 
with high redshifts.
Therefore, the soft sources show the break due to 
the cosmological evolutionary effects 
and mixture of X-ray populations. 

4. The resolved CXRB flux densities are measured
from the ChaMP and the ChaMP+CDFs number counts in multiple 
energy bands. 
We present upper and lower limits of the resolved CXRB 
by estimating with and without bright target sources.

5. Excluding target sources, the total CXRB flux densities 
in units of $erg$ $cm^{-2}$ $sec^{-1}$ $deg^{-2}$ are 
$0.81^{+0.03}_{-0.03}\times 10^{-11}$ (0.5-2 keV),
$1.74^{+0.17}_{-0.17}\times 10^{-11}$ (2-8 keV), and
$0.48^{+0.02}_{-0.02}\times 10^{-11}$ (1-2 keV), respectively.
Including target sources, the total CXRB flux densities 
in units of $erg$ $cm^{-2}$ $sec^{-1}$ $deg^{-2}$ are
$0.86^{+0.03}_{-0.03}\times 10^{-11}$ (0.5-2 keV),
$1.84^{+0.18}_{-0.18}\times 10^{-11}$ (2-8 keV), and
$0.51^{+0.02}_{-0.02}\times 10^{-11}$ (1-2 keV), respectively.

6. When the total CXRB estimated from this study is assumed in each band,  
excluding target sources, the resolved CXRB fractions are 
$78.1^{+1.2}_{-1.2}\%$ (0.5-2 keV),
$80.5^{+1.7}_{-1.7}\%$ (2-8 keV),
$78.5^{+1.2}_{-1.2}\%$ (1-2 keV), respectively. 
Including target sources, the resolved CXRB fractions are
$79.3^{+1.2}_{-1.2}\%$ (0.5-2 keV),
$81.5^{+3.8}_{-3.8}\%$ (2-8 keV), and
$79.8^{+1.2}_{-1.2}\%$ (1-2 keV), respectively.

We gratefully acknowledge support for this project under NASA CXC
archival research grant AR4-5017X and AR6-7020X .  PJG, DWK,
HT, and BJW also acknowledge support through NASA Contract
NAS8-03060 (CXC).
MGL is in part supported by the KOSEF grant
(R01-2004-000-10490-0).


\begin{center}
\begin{deluxetable}{ccr}
\tablecaption{Definition of Energy Bands \label{tbl-energy-def}}
\tablewidth{0pt}
\tabletypesize{\normalsize}
\tablecolumns{2}\tablehead{\colhead{Range}&
\colhead{Band}&
\colhead{Definition}}
\startdata
Broad  & B    &    0.3-8 keV \\
       & Bc   &    0.5-8 keV \\
\hline
Soft   & S    &    0.3-2.5 keV \\
       & Sc   &    0.5-2 keV \\
       & Ss\tablenotemark{a}   &      1-2 keV \\
\hline
Hard   & H    &    2.5-8 keV \\
       & Hc   &    2-8 keV \\
       & He\tablenotemark{b}   &    2-10 keV \\
\enddata
\tablenotetext{a}{The Ss (2-10 keV) band was used only for 
estimating the CXRB flux density (see \S 6).}
\tablenotetext{b}{The He (2-10 keV) band was not used in this
study; however, it is referred for previous studies.}
\end{deluxetable}
\end{center}
\clearpage

\begin{center}
\begin{deluxetable}{lc|crrrr}
\tablecaption{Statistical Properties of X-ray Point Sources \label{tbl-prop}}
\tablewidth{0pt}
\tabletypesize{\normalsize}
\tablecolumns{7}\tablehead{
\colhead{}&
\colhead{Band}&
\colhead{Number}&
\colhead{Min}&
\colhead{Max}&
\colhead{Median}&
\colhead{Mean}\\
\colhead{}&
\colhead{(1)}&
\colhead{(2)}&
\colhead{(3)}&
\colhead{(4)}&
\colhead{(5)}&
\colhead{(6)}}
\startdata
Count\tablenotemark{a} &     B &        5515 &         5.42 &     40535.59 &        22.57 &        69.53 \\ 
 &     S &        4864 &         5.42 &     38117.52 &        19.24 &        61.50 \\ 
 &     H &        2575 &         5.41 &     11604.93 &        12.73 &        28.63 \\ 
 &    Bc &        5229 &         6.46 &     39760.98 &        23.46 &        70.52 \\ 
 &    Sc &        4554 &         5.41 &     36010.96 &        18.24 &        57.59 \\ 
 &    Hc &        3078 &         5.42 &     13624.92 &        13.72 &        31.63 \\ 
\hline
Flux\tablenotemark{b} &     B &        5515 &         0.63 &      7175.62 &         9.09 &        25.97 \\ 
 &     S &        4864 &         0.33 &      3286.49 &         4.36 &        12.78 \\ 
 &     H &        2575 &         1.27 &      6690.72 &         8.61 &        21.40 \\ 
 &    Bc &        5229 &         0.69 &      6767.74 &         9.04 &        25.38 \\ 
 &    Sc &        4554 &         0.26 &      2395.21 &         3.21 &         9.32 \\ 
 &    Hc &        3078 &         1.17 &      7112.31 &         8.87 &        21.88 \\ 
\enddata
\tablecomments{
Col. (1): X-ray energy band (see Table \ref{tbl-energy-def}).
Col. (2): number of sources.
Col. (3): minimum value of the sample.
Col. (4): maximum value of the sample.
Col. (5): median value of the sample.
Col. (6): mean value of the sample.
}
\tablenotetext{a}{Source net counts.}
\tablenotetext{b}{Source flux with $\Gamma_{ph}=1.4$ 
in units of $10^{-15}$ $erg$ $cm^{-2}$ $sec^{-1}$.}
\end{deluxetable}
\end{center}
\clearpage

\begin{center}
\begin{deluxetable}{cccrrrr}
\tablecaption{List of the Best Fit Parameters without Target Objects \label{tbl-fit_diff_nt}}
\tablewidth{0pt}
\tabletypesize{\normalsize}
\tablehead{
\colhead{DATA}&
\colhead{$\Gamma_{ph}$}&
\colhead{Band}&
\colhead{$K$}& 
\colhead{$\gamma_{1}$}&
\colhead{$\gamma_{2}$}&
\colhead{$S_{b}$}\\ 
\colhead{(1)}&
\colhead{(2)}&
\colhead{(3)}&
\colhead{(4)}&
\colhead{(5)}&
\colhead{(6)}&
\colhead{(7)}
}
\startdata
ChaMP & 1.4 &       S & 769$^{+  14}_{ -14}$ & 1.57$^{+0.01}_{-0.01}$ &         2.41$^{+0.05}_{-0.05}$ &  9.9$^{+ 0.7}_{-1.6}$ \\ 
      &     &      H &1828$^{+  48}_{ -43}$ & 1.81$^{+0.01}_{-0.01}$ &         2.58$^{+0.05}_{-0.05}$ & 14.2$^{+ 0.9}_{-1.1}$\\ 
      &     &      B & 1614$^{+  28}_{ -43}$ & 1.65$^{+0.01}_{-0.01}$ &         2.44$^{+0.06}_{-0.05}$ & 25.0$^{+ 1.9}_{-1.9}$\\ 
\hline
      & 1.7 &       S &  783$^{+  15}_{ -15}$ & 1.58$^{+0.01}_{-0.01}$ &    2.42$^{+0.05}_{-0.05}$ & 10.5$^{+ 0.8}_{-0.8}$ \\ 
      &     &   H & 1774$^{+  44}_{ -41}$ & 1.80$^{+0.01}_{-0.01}$ &    2.58$^{+0.05}_{-0.05}$ & 13.5$^{+ 0.9}_{-0.9}$\\ 
      &     &       B & 1505$^{+  25}_{ -41}$ & 1.65$^{+0.01}_{-0.01}$ &    2.45$^{+0.06}_{-0.05}$ & 21.9$^{+ 1.7}_{-1.7}$\\ 
\hline
      & 1.4   &       Sc & 607$^{+  12}_{ -12}$ & 1.54$^{+0.02}_{-0.02}$ &         2.36$^{+0.05}_{-0.05}$ &  6.8$^{+ 0.5}_{-0.5}$ \\ 
      & &       Hc &2040$^{+  50}_{ -50}$ & 1.82$^{+0.01}_{-0.01}$ &         2.65$^{+0.07}_{-0.07}$ & 19.2$^{+ 6.3}_{-1.8}$\\ 
      &    &       Bc & 1557$^{+  28}_{ -50}$ & 1.64$^{+0.01}_{-0.01}$ &         2.48$^{+0.05}_{-0.05}$ & 22.9$^{+ 1.6}_{-1.6}$\\ 
\hline
      & 1.7 &       Sc &  612$^{+  12}_{ -12}$ & 1.53$^{+0.02}_{-0.02}$ &    2.36$^{+0.05}_{-0.04}$ &  6.7$^{+ 0.5}_{-0.5}$ \\ 
      &     &       Hc & 1932$^{+  46}_{ -48}$ & 1.82$^{+0.01}_{-0.01}$ &    2.64$^{+0.07}_{-0.07}$ & 17.8$^{+ 4.4}_{-1.7}$\\ 
      &    &       Bc & 1407$^{+  25}_{ -48}$ & 1.64$^{+0.01}_{-0.01}$ &    2.48$^{+0.05}_{-0.05}$ & 19.2$^{+ 1.3}_{-1.4}$\\ 
\hline
ChaMP+ & 1.4 &       Sc & 571$^{+  11}_{ -11}$ & 1.49$^{+0.02}_{-0.02}$ &
  2.36$^{+0.05}_{-0.05}$ &  6.5$^{+ 0.4}_{-0.4}$ \\
CDFs   &     &       Hc &1258$^{+  29}_{ -29}$ & 1.58$^{+0.01}_{-0.01}$ &
  2.59$^{+0.06}_{-0.05}$ & 14.4$^{+ 0.9}_{-0.9}$\\
\hline
\enddata
\tablecomments{
Col. (1): used data set.
Col. (2): assumed photon index.
Col. (3): X-ray energy band (see Table \ref{tbl-energy-def}).
Col. (4): normalization constant. 
Col. (5): faint power index of a broken power law.
Col. (6): bright power index of a broken power law.
Col. (7): break flux in units of $10^{-15}$ $erg~cm^{-2}~sec^{-1}$.
}
\end{deluxetable}
\end{center}
\clearpage

\begin{center}
\begin{deluxetable}{lrrcrrrrrrrrrcr}
\rotate
\tablecaption{List of Fitting Parameters of Other Studies, ChaMP, and ChaMP+CDFs \label{tbl-other}}
\tablewidth{0pt}
\tabletypesize{\tiny}
\tablehead{
\colhead{Data}&
\colhead{Area}&
\colhead{Band}&
\colhead{$\Gamma_{ph}$}&
\colhead{Number}&
\colhead{$K$}&
\colhead{$S_{ref}$}&
\colhead{$\gamma_{1}$}&
\colhead{$\gamma_{2}$}&
\colhead{$S_{b}$}&
\colhead{$f_{min}$}&
\colhead{$f_{max}$}&
\colhead{FS}&
\colhead{FM}&
\colhead{Reference}\\
\colhead{(1)}& 
\colhead{(2)}& 
\colhead{(3)}& 
\colhead{(4)}& 
\colhead{(5)}& 
\colhead{(6)}& 
\colhead{(7)}& 
\colhead{(8)}& 
\colhead{(9)}& 
\colhead{(10)}& 
\colhead{(11)}& 
\colhead{(12)}& 
\colhead{(13)}& 
\colhead{(14)}& 
\colhead{(15)} 
}
\startdata
SSA13 & 0.03 & $0.5-2$  & 1.4 & 22  & $185$    &  7   & $0.7\pm0.2$     &  &  & 0.23 & 7 & C & S & \cite{mus00} \\ 
      &      & $2-10$   & 1.2 & 15  & $170$    & 20   & $1.05\pm0.35$   &  &  & 2.5  & 20  & C & S &              \\ 
\hline
HELLAS2XMM & 3   & $0.5-2$  & 1.7 & 1022 &         & 10   & $1.1\sim1.7$    & $2.2^{+0.06}_{-0.09}$ & $5\sim6.5$ & 0.59 & 500 & D & B & \cite{bal02} \\
           &     & $0.5-2$  & 1.7 & 1022 & $80.8^{+6.4}_{-5.2}$    &    & $0.93\pm0.05$          &   &   & 0.59 & 500 & C & S &           \\
           &     & $2-10$   & 1.7 & 495  & $229^{+29.3}_{-19.6}$   &    & $1.34^{+0.11}_{-0.10}$ &   &   & 2.8 & 6000 & C & S &           \\
           &     & $5-10$   & 1.7 & 100  & $175.2^{+56.3}_{-36.2}$ &    & $1.54^{+0.25}_{-0.19}$ &   &   & 6.2 & 1000 & C & S &           \\
\hline
CDFs+   & 0.25 & $2-8$    & 1.2 & 373 & $32\pm2$   &  10    & $1.63\pm0.05$   &               & 12 & 0.2 & 100 & D & D & \cite{cow02} \\
SSA13/SSA22 &      &          &     &     & $39\pm5$   &  10    &                 & $2.57\pm0.22$ & 12 & 0.2 & 100 & D & D &              \\ 
\hline
CDF-S & & $0.5-2$  & 1.4 & 346 & $380\pm80$     &      & $1.63\pm0.13$ &     & $\sim13$ & 0.06 & 50 & D & D & \cite{ros02} \\ 
      & & $2-10$   & 1.4 & 251 & $1300\pm100$   &      & $1.61\pm0.10$ &     & $\sim8$  & 0.45 & 90 & D & D &              \\
      & & $5-10$   & 1.4 & 110 & $940\pm100$    &      & $1.35\pm0.15$ &     &          & 1    & 40 & D & D &              \\             
\hline 
SEXSI & 2.1 & $2-10$   &  1.5  &  478  & $\sim43.65^{+2.1}_{-2.0}$  & 10  & $1.41\pm0.17$ &               & $\sim11$ & 1 & 100 & D & D & H03 \\
      &     &          &       &       & $\sim46.8\pm2.1$           & 10  &               & $2.46\pm0.08$ & $\sim11$ & 1 & 100 & D & D &              \\
\hline
ELAIS & 0.17 & $0.5-2$  & 1.7  & 182   &  630     &      & $1.72\pm0.09$   &               &            & 0.57 & 26 & D & S & \cite{man03} \\
      &      & $2-8$    & 1.7  & 124   & 3548     &      & $2.07\pm0.15$   &               &            & 2.7  & 63 & D & S &              \\
      &      & $0.5-8$  & 1.7  & 225   & 1258     &      & $1.70\pm0.08$   &               &            & 1.4  & 70 & D & S &              \\
\hline
BMW\tablenotemark{a}    & 91.64 & $0.5-2$  & 1.4 & 4786 & $6150^{+1800}_{-1650}$ &   & $1.82^{+0.07}_{-0.09}$  & $0.60^{+0.02}_{-0.03}$ & $14.8^{+1.27}_{-1.31}$ & 0.02 & 10000 & D & N$\dagger$ & M03 \\
$ASCA$\tablenotemark{b} & 73.71 & $2-10$   & 1.4 & 1026 & $5300^{+2850}_{-1400}$ &   & $1.57^{+0.10}_{-0.08}$  & $0.44^{+0.12}_{-0.13}$ & $4.5^{+3.7}_{-1.7}$ & 0.21    & 8000 & D & N$\dagger$ &               \\
\hline
CDF-N+CDF-S& 0.2 & $0.5-2$  & 1.4 & 724 & $3039^{+88}_{-108}$  &  & $0.55\pm0.03$ &     &          & 0.02 & 83.73 & C & S & B04 \\
           &     & $2-8$    & 1.4 & 520 & $7403^{+125}_{-599}$ &  & $0.56\pm0.14$ &     &          & 0.19 & 140.80& C & S &              \\
\hline
ChaMP & 1.1 & $0.5-2$ & 1.7 & 707 & $2030\pm210$    &  1   & $1.40\pm0.30$   & $2.2\pm0.20$  & $6\pm2$ & 0.6 & 100 & D & B & KD04 \\  
      &     & $2-8$   & 1.4 & 236 & $3160\pm250$    &      &                 & $2.10\pm0.10$ &         &   4 & 400 & C & S &               \\
\hline
CLASXS & 0.4 & $0.5-2$ & 1.4   & 310   & $12.49\pm0.02$  &     & $1.7\pm0.2$     &               & $\sim10$     & 0.5 & 35 & D & D & \cite{yan04}  \\  
       &     & $0.5-2$ & 1.4   & 310   & $78.81$         &     &                 & $2.5 (fixed)$ & $\sim10$     & 0.5 & 35 & D & D &               \\  
       &     & $2-8$   & 1.4   & 235   & $38.1\pm0.2$    &     & $1.65\pm0.4$    &               & $10\sim30$   & 3   & 90 & D & D &               \\  
       &     & $2-8$   & 1.4   & 235   & $45.60\pm0.5$   &     &                 & $2.4\pm0.6$   & $10\sim30$   & 3   & 90 & D & D &               \\  
\hline
XMM/2dF& 2   & $0.5-2$ &       & 432   &                 &     &                 &               &          & 2.7 & $\sim500$ &   &   & \cite{bas05} \\
       &     & $0.5-8$ &       & 462   &                 &     & $1.8\pm0.2$     &               & $\sim60$ & 6.0 & $\sim700$ & D & D &              \\ 
       &     &         &       &       &                 &     &                 & $2.3\pm0.1$   & $\sim60$ & 6.0 & $\sim700$ & D & D &              \\ 
\hline 
XMM-LSS& 3.4 & $0.5-2$ & 1.7 & 1028 & $384.2$         & 1  & $1.42^{+0.14}_{-0.15}$ &                        & $10.6^{+3.0}_{-2.2}$ & $\sim1$ & 700 & D & D & \cite{chi05} \\ 
       &     & $0.5-2$ & 1.7 & 1028 & $6515$          & 1  &                        & $2.62^{+0.25}_{-0.22}$ & $10.6^{+3.0}_{-2.2}$ & $\sim1$ & 700 & D & D &              \\ 
       &     & $2-10$  & 1.7 & 328  &                 & 1  & $1.53^{+0.51}_{-1.16}$ &                        & $21.4^{+8.1}_{-5.4}$ & $\sim7$ & 500 & D & D &              \\
       &     & $2-10$  & 1.7 & 328  & $4.5\times10^{4}$ & 1  &                      & $2.91^{+0.45}_{-0.30}$ & $21.4^{+8.1}_{-5.4}$ & $\sim7$ & 500 & D & D &              \\
\hline
ChaMP& 9.6 & $0.5-2$   & 1.4 & 4554    & 607$^{+12}_{-12}$    & 1 & 1.54$^{+0.02}_{-0.02}$ & 2.36$^{+0.05}_{-0.05}$ & 6.8$^{+0.5}_{-0.5}$   & 0.26 &2395.21 & D & B  & this study \\ 
     &     & $2-8$     & 1.4 & 3078    & 2040$^{+50}_{-50}$ & 1 & 1.82$^{+0.01}_{-0.01}$ & 2.65$^{+0.07}_{-0.07}$ & 19.2$^{+6.3}_{-1.8}$ & 1.17 &7112.31 & D & B & \\
     &     & $0.5-8$   & 1.4 & 5229    & 1557$^{+28}_{-50}$ & 1 & 1.64$^{+0.01}_{-0.01}$ & 2.48$^{+0.05}_{-0.05}$ & 22.9$^{+1.6}_{-1.6}$  & 0.69 &6767.74 & D & B & \\
     &     & $0.3-2.5$ & 1.4 & 4864    & 769$^{+14}_{-14}$  & 1 & 1.57$^{+0.01}_{-0.01}$ & 2.41$^{+0.05}_{-0.05}$ & 9.9$^{+0.7}_{-1.6}$   & 0.33 &3286.49 & D & B & \\
     &     & $2.5-8$   & 1.4 & 2575    & 1828$^{+48}_{-43}$ & 1 & 1.81$^{+0.01}_{-0.01}$ & 2.58$^{+0.05}_{-0.05}$ & 14.2$^{+0.9}_{-1.1}$  & 1.27 &6690.72 & D & B & \\
     &     & $0.3-8$   & 1.4 & 5515    & 1614$^{+28}_{-43}$ & 1 & 1.65$^{+0.01}_{-0.01}$ & 2.44$^{+0.06}_{-0.05}$ & 25.0$^{+1.9}_{-1.9}$  & 0.63 &7175.62 & D & B & \\
\hline
ChaMP + CDFs & 9.8 & $0.5-2$ & 1.4  & $4554+724$   &  571$^{+11}_{-11}$   & 1 & 1.49$^{+0.02}_{-0.02}$ & 2.36$^{+0.05}_{-0.05}$  & 6.5$^{+0.4}_{-0.4}$  & 0.02  &2395.21 & D & B  & this study \\
             &     & $2-8$   & 1.4  & $3078+520$   & 1258$^{+29}_{-29}$ & 1 & 1.58$^{+0.01}_{-0.01}$ & 2.59$^{+0.06}_{-0.05}$  & 14.4$^{+0.9}_{-0.9}$ & 0.19 &7112.31 & D & B  &  \\
\enddata
\tablecomments{
Col. (1): used data. $^{a}$BMW (Brera Multiscale Wavelet)-HRI, 
HELLAS2XMM, BMW-CDFS, and BMW-HDF. $^{b}$ASCA-HSS, HELLAS2XMM, BMW-CDFS, and BMW-HDF.
Col. (2): sky coverage of the sample in units of $deg^{2}$. 
Col. (3): X-ray energy band in units of keV.
Col. (4): assumed photon index.
Col. (5): source numbers in the sample.
Col. (6): normalization constant. 
Col. (7): normalization flux in units of $10^{-15}erg~cm^{-2}~sec^{-1}$.
Col. (8): faint power law index.
Col. (9): bright power law index. 
Col. (10): break flux in units of $10^{-15}erg~cm^{-2}~sec^{-1}$.
Col. (11): faint flux limit of the sample in units of $10^{-15}erg~cm^{-2}~sec^{-1}$.
Col. (12): bright flux limit of the sample in units of $10^{-15}erg~cm^{-2}~sec^{-1}$.
Col. (13): fitting domain. C: cumulative number count, D: differential number count.
Col. (14): fitting formula. S: single power law, B: broken power law, D: double power law,
N: nonlinear formula, see equation (5) in text. 
Col. (15): reference.
}
\end{deluxetable}
\end{center}
\clearpage

\begin{center}
\begin{deluxetable}{cc|rrrrr|rrr}
\tablecaption{List of the Best Fit Parameters of 
the Soft and Hard Sources\label{tbl-fit_diff_nt_hr}}
\tablewidth{0pt}
\tabletypesize{\small}
\tablehead{
\colhead{}&
\colhead{}{}&
\multicolumn{5}{c}{Soft Source (HR$<0$)}&
\multicolumn{3}{c}{Hard Source (HR$>0$)}\\
\colhead{$\Gamma_{ph}$}&
\colhead{Band}&
\colhead{N}&
\colhead{$K$}& 
\colhead{$\gamma_{1}$}&
\colhead{$\gamma_{2}$}&
\colhead{$S_{b}$}&
\colhead{N}&
\colhead{$K$}&
\colhead{$\gamma$}\\
\colhead{(1)}&
\colhead{(2)}&
\colhead{(3)}&
\colhead{(4)}&
\colhead{(5)}&
\colhead{(6)}&
\colhead{(7)}&
\colhead{(8)}&
\colhead{(9)}&
\colhead{(10)}
}
\startdata
       1.4 &      S &   4289 &  580$^{+  12}_{ -12}$ & 1.43$^{+0.02}_{-0.02}$ &         2.35$^{+0.04}_{-0.04}$ &  8.5$^{+ 0.5}_{-0.5}$        &    575 &  217$^{+  22}_{ -20}$ & 2.45$^{+0.08}_{-0.07}$  \\ 
            &       H &   1787 &  896$^{+  31}_{ -31}$ & 1.63$^{+0.02}_{-0.02}$ &         2.58$^{+0.08}_{-0.07}$ & 13.5$^{+ 1.1}_{-1.3}$        &    787 & 1614$^{+ 209}_{-184}$ & 2.35$^{+0.05}_{-0.05}$  \\ 
            &       B &   4427 &  900$^{+  18}_{ -31}$ & 1.46$^{+0.01}_{-0.01}$ &         2.35$^{+0.04}_{-0.04}$ & 20.6$^{+ 1.4}_{-1.3}$        &   1088 & 1195$^{+ 106}_{ -98}$ & 2.34$^{+0.04}_{-0.04}$  \\ 
\hline
    1.7 &       S &   4289 &  588$^{+  12}_{ -12}$ & 1.44$^{+0.02}_{-0.02}$ &    2.34$^{+0.04}_{-0.04}$ &  8.7$^{+ 0.6}_{-0.5}$     &    575 &  237$^{+  23}_{ -22}$ & 2.46$^{+0.08}_{-0.07}$  \\ 
        &       H &   1787 &  898$^{+  31}_{ -31}$ & 1.64$^{+0.02}_{-0.02}$ &    2.59$^{+0.07}_{-0.07}$ & 13.1$^{+ 0.9}_{-1.3}$    &    787 & 1508$^{+ 192}_{-170}$ & 2.35$^{+0.05}_{-0.05}$  \\ 
        &       B &   4427 &  924$^{+  17}_{ -31}$ & 1.51$^{+0.01}_{-0.02}$ &    2.42$^{+0.06}_{-0.09}$ & 21.0$^{+ 1.5}_{-4.0}$    &   1088 & 1037$^{+  86}_{ -80}$ & 2.35$^{+0.04}_{-0.04}$  \\ 
\hline
       1.4 &      Sc &   4149 &  521$^{+  11}_{ -11}$ & 1.47$^{+0.02}_{-0.02}$ &         2.35$^{+0.05}_{-0.05}$ &  6.7$^{+ 0.5}_{-0.4}$        &    405 &  101$^{+  11}_{ -10}$ & 2.47$^{+0.10}_{-0.10}$  \\ 
            &       Hc &   2185 & 1129$^{+  33}_{ -33}$ & 1.69$^{+0.01}_{-0.01}$ &         2.57$^{+0.07}_{-0.07}$ & 16.4$^{+ 4.4}_{-1.5}$        &    893 & 1509$^{+ 182}_{-163}$ & 2.30$^{+0.05}_{-0.04}$  \\ 
            &       Bc &   4235 &  916$^{+  18}_{ -33}$ & 1.49$^{+0.01}_{-0.01}$ &         2.45$^{+0.05}_{-0.05}$ & 21.8$^{+ 1.4}_{-1.4}$        &    994 & 1160$^{+ 116}_{-105}$ & 2.37$^{+0.05}_{-0.04}$  \\ 
\hline
    1.7 &       Sc &   4149 &  525$^{+  11}_{ -11}$ & 1.47$^{+0.02}_{-0.02}$ &    2.35$^{+0.05}_{-0.05}$ &  6.8$^{+ 0.5}_{-0.5}$     &    405 &  100$^{+  10}_{ -10}$ & 2.44$^{+0.10}_{-0.09}$  \\ 
        &       Hc &   2185 & 1060$^{+  31}_{ -31}$ & 1.68$^{+0.01}_{-0.01}$ &    2.58$^{+0.07}_{-0.07}$ & 15.4$^{+ 2.0}_{-1.4}$    &    893 & 1378$^{+ 164}_{-147}$ & 2.30$^{+0.05}_{-0.05}$  \\ 
        &       Bc &   4235 &  842$^{+  17}_{ -31}$ & 1.48$^{+0.01}_{-0.01}$ &    2.44$^{+0.05}_{-0.05}$ & 18.1$^{+ 1.2}_{-3.0}$    &    994 &  946$^{+  87}_{ -81}$ & 2.38$^{+0.05}_{-0.04}$  \\ 
\hline
\enddata
\tablecomments{
Col. (1): assumed photon index. 
Col. (2): X-ray energy band (see Table \ref{tbl-energy-def}).
Col. (3): number of sources with a hardness ratio of HR$<0$.
Col. (4): normalization constant of a broken power law.
Col. (5): faint power law index of a broken power law.
Col. (6): bright power law index of a broken power law.
Col. (7): break flux in units of $10^{-15}$ $erg~cm^{-2}~sec^{-1}$.
Col. (8): number of sources with a hardness ratio of HR$>0$.
Col. (9): normalization constant of a single power law.
Col. (10): power law index of a single power law.
}
\end{deluxetable}
\end{center}
\clearpage

\begin{center}
\begin{deluxetable}{cccc}
\tablecaption{Break Factor\label{tbl-break_factor}}
\tablewidth{0pt}
\tabletypesize{\normalsize}
\tablecolumns{4}\tablehead{\colhead{$\Gamma_{ph}$}&
\colhead{Band}&
\colhead{Total}&
\colhead{HR$<0$}\\
\colhead{(1)}&
\colhead{(2)}&
\colhead{(3)}&
\colhead{(4)}}
\startdata
1.4 & B  &   0.19 &   0.23 \\
    & S  &   0.21 &   0.24 \\
    & H  &   0.18 &   0.23 \\
    & Bc &   0.20 &   0.24 \\
    & Sc &   0.21 &   0.23 \\
    & Hc &   0.18 &   0.21 \\
\hline
1.7 & B  &   0.20 &   0.21 \\
    & S  &   0.21 &   0.24 \\
    & H  &   0.18 &   0.22 \\
    & Bc &   0.20 &   0.25 \\
    & Sc &   0.21 &   0.23 \\
    & Hc &   0.18 &   0.21 \\
\enddata
\tablecomments{
Col. (1): assumed photon index. 
Col. (2): X-ray energy band (see Table \ref{tbl-energy-def}).
Col. (3): break factor for the total sample.
Col. (4): break factor for the soft sample (HR$<0$).}
\end{deluxetable}
\end{center}
\clearpage

\begin{center}
\begin{deluxetable}{cccrrrr}
\tablecaption{List of the Best Fit Parameters including Target Objects \label{tbl-fit_diff_yt}}
\tablewidth{0pt}
\tabletypesize{\normalsize}
\tablehead{
\colhead{DATA}&
\colhead{$\Gamma_{ph}$}&
\colhead{Band}&
\colhead{$K$}& 
\colhead{$\gamma_{1}$}&
\colhead{$\gamma_{2}$}&
\colhead{$S_{b}$}\\ 
\colhead{(1)}&
\colhead{(2)}&
\colhead{(3)}&
\colhead{(4)}&
\colhead{(5)}&
\colhead{(6)}&
\colhead{(7)}
}
\startdata
ChaMP & 1.4 &       S & 753$^{+  15}_{ -15}$ & 1.54$^{+0.02}_{-0.01}$ &         2.31$^{+0.04}_{-0.04}$ &  8.6$^{+ 0.7}_{-0.6}$ \\ 
      &     &      H &1856$^{+  53}_{ -48}$ & 1.81$^{+0.01}_{-0.01}$ &         2.48$^{+0.05}_{-0.06}$ & 14.0$^{+ 0.9}_{-1.6}$\\ 
      &     &      B & 1550$^{+  28}_{ -48}$ & 1.62$^{+0.01}_{-0.01}$ &         2.31$^{+0.04}_{-0.04}$ & 21.0$^{+ 4.7}_{-1.6}$\\ 
\hline
      & 1.7 &       S &  766$^{+  15}_{ -15}$ & 1.55$^{+0.01}_{-0.01}$ &    2.31$^{+0.04}_{-0.04}$ &  9.0$^{+ 1.7}_{-0.6}$ \\ 
      &     &   H & 1825$^{+  50}_{ -41}$ & 1.82$^{+0.01}_{-0.01}$ &    2.48$^{+0.04}_{-0.05}$ & 13.5$^{+ 0.7}_{-1.4}$\\ 
      &     &       B & 1469$^{+  26}_{ -41}$ & 1.63$^{+0.01}_{-0.01}$ &    2.34$^{+0.04}_{-0.04}$ & 19.3$^{+ 3.6}_{-1.4}$\\ 
\hline
      & 1.4   &       Sc & 610$^{+  12}_{ -12}$ & 1.54$^{+0.02}_{-0.02}$ &         2.30$^{+0.05}_{-0.04}$ &  6.6$^{+ 0.5}_{-0.5}$ \\ 
      & &       Hc &2038$^{+  50}_{ -50}$ & 1.82$^{+0.01}_{-0.01}$ &         2.54$^{+0.07}_{-0.06}$ & 18.5$^{+ 2.5}_{-2.1}$\\ 
      &    &       Bc & 1561$^{+  27}_{ -50}$ & 1.64$^{+0.01}_{-0.01}$ &         2.40$^{+0.05}_{-0.05}$ & 22.1$^{+ 1.7}_{-1.7}$\\ 
\hline
      & 1.7 &       Sc &  615$^{+  12}_{ -12}$ & 1.53$^{+0.02}_{-0.02}$ &    2.29$^{+0.04}_{-0.04}$ &  6.5$^{+ 0.5}_{-0.5}$ \\ 
      &     &       Hc & 1930$^{+  47}_{ -47}$ & 1.82$^{+0.01}_{-0.01}$ &    2.53$^{+0.07}_{-0.07}$ & 17.1$^{+ 2.0}_{-2.2}$\\ 
      &    &       Bc & 1410$^{+  25}_{ -47}$ & 1.64$^{+0.01}_{-0.01}$ &    2.39$^{+0.05}_{-0.05}$ & 18.4$^{+ 1.5}_{-1.4}$\\ 
\hline
ChaMP+ & 1.4 &       Sc & 574$^{+  12}_{ -12}$ & 1.49$^{+0.02}_{-0.02}$ &         2.29$^{+0.05}_{-0.10}$ &  6.2$^{+ 0.5}_{-1.4}$ \\ 
CDFs   &     &       Hc &1240$^{+  55}_{ -55}$ & 1.55$^{+0.02}_{-0.02}$ &         2.44$^{+0.04}_{-0.04}$ & 11.9$^{+ 0.8}_{-0.7}$\\ 
\hline
\enddata
\tablecomments{
Col. (1): used data set.
Col. (2): assumed photon index.
Col. (3): X-ray energy band (see Table \ref{tbl-energy-def}).
Col. (4): normalization constant. 
Col. (5): faint power law index of a broken power law.
Col. (6): bright power law index of a broken power law.
Col. (7): break flux in units of $10^{-15}$ $erg~cm^{-2}~sec^{-1}$.
}
\end{deluxetable}
\end{center}
\clearpage

\begin{center}
\begin{deluxetable}{ccccccccc}
\tablecaption{The Resolved Cosmic X-ray Background Flux Density  \label{tbl-cxrb_resol}}
\tablewidth{0pt}
\tabletypesize{\footnotesize}
\tablecolumns{9}\tablehead{
\colhead{Data}&
\colhead{Band}&
\colhead{$f_{min}$}& 
\colhead{$f_{max}$}&
\colhead{$CXRB_{total}$}&
\colhead{$CXRB_{nt}$}&
\colhead{$Fraction_{nt} [\%]$}&
\colhead{$CXRB_{yt}$}&
\colhead{$Fraction_{yt} [\%]$}\\
\colhead{(1)}&
\colhead{(2)}&
\colhead{(3)}&
\colhead{(4)}&
\colhead{(5)}&
\colhead{(6)}&
\colhead{(7)}&
\colhead{(8)}&
\colhead{(9)}}
\startdata
ChaMP &  B & $ 0.63 $ &    $ 0.72 $ &    $ 2.70 \pm {0.12}       $ &    $ 2.10^{+0.04}_{-0.04}  $ &    $ 77.6^{+ 1.3}_{-1.3}  $ &    $ 2.29^{+0.04}_{-0.04}  $ &    $ 84.9^{+ 1.5}_{-1.5}  $ \\ 
&  S & $ 0.33 $ &    $ 0.33 $ &    $ 1.10 \pm {0.05}       $ &    $ 0.82^{+0.02}_{-0.02}  $ &    $ 74.7^{+ 1.4}_{-1.4}  $ &    $ 0.89^{+0.02}_{-0.02}  $ &    $ 81.5^{+ 1.6}_{-1.6}  $ \\ 
&  H & $ 1.27 $ &    $ 0.67 $ &    $ 1.59 \pm {0.10}       $ &    $ 1.10^{+0.03}_{-0.03}  $ &    $ 69.4^{+ 1.8}_{-1.6}  $ &    $ 1.19^{+0.03}_{-0.03}  $ &    $ 74.5^{+ 2.1}_{-1.9}  $ \\ 
& Bc & $ 0.69 $ &    $ 0.68 $ &    $ 2.54 \pm {0.12}       $ &    $ 1.88^{+0.03}_{-0.03}  $ &    $ 74.0^{+ 1.3}_{-1.3}  $ &    $ 2.01^{+0.04}_{-0.04}  $ &    $ 79.2^{+ 1.4}_{-1.4}  $ \\ 
& Sc & $ 0.26 $ &    $ 0.24 $ &    $ 0.75 \pm {0.04}       $ &    $ 0.60^{+0.01}_{-0.01}  $ &    $ 80.1^{+ 1.6}_{-1.6}  $ &    $ 0.65^{+0.01}_{-0.01}  $ &    $ 86.3^{+ 1.7}_{-1.7}  $ \\ 
& Hc & $ 1.17 $ &    $ 0.71 $ &    $ 1.79 \pm {0.11}       $ &    $ 1.28^{+0.03}_{-0.03}  $ &    $ 71.6^{+ 1.7}_{-1.7}  $ &    $ 1.36^{+0.03}_{-0.03}  $ &    $ 76.1^{+ 1.8}_{-1.8}  $ \\ 
\hline
ChaMP+ & Sc & $ 0.02 $ &    $ 0.24 $ &    $ 0.75 \pm {0.04}       $ &    $ 0.63^{+0.01}_{-0.01}  $ &    $ 84.4^{+ 1.6}_{-1.6}  $ &    $ 0.68^{+0.01}_{-0.01}  $ &    $ 90.7^{+ 1.9}_{-1.9}  $ \\ 
CDFs   & Hc & $ 0.20 $ &    $ 0.71 $ &    $ 1.79 \pm {0.11}       $ &    $ 1.40^{+0.03}_{-0.03}  $ &    $ 78.1^{+ 1.8}_{-1.8}  $ &    $ 1.50^{+0.07}_{-0.07}  $ &    $ 84.0^{+ 3.7}_{-3.7}  $ \\ 
\enddata
\tablecomments{
Col. (1): used data set.
Col. (2): X-ray energy band (see Table \ref{tbl-energy-def}).
Col. (3)-(4): faint and bright flux limits of the data 
in units of $10^{-15}~erg~cm^{-2}~sec^{-1}$ and $10^{-11}~erg~cm^{-2}~sec^{-1}$, respectively. 
Col. (5): total CXRB flux density in units 
of $10^{-11}~erg~cm^{-2}~sec^{-1}~deg^{-2}$. 
The total CXRB flux densities in the Sc and Hc bands are from B04.
The total CXRB flux density in the Bc band is the sum of those in the Sc and Hc bands.
The total CXRB flux densities in the B, S, and H bands are rescaled 
from those in the Bc, Sc, and Hc bands by assuming $\Gamma_{ph}=1.1$.
Col. (6): the resolved CXRB flux density  
without target sources in units of $10^{-11}~erg~cm^{-2}~sec^{-1}~deg^{-2}$.
Col. (7): the percentage of the resolved CXRB excluding target sources.
Col. (8): the resolved CXRB flux density  
with target sources in units of $10^{-11}~erg~cm^{-2}~sec^{-1}~deg^{-2}$.
Col. (9): the percentage of the resolved CXRB including target sources.
}
\end{deluxetable}
\end{center}
\clearpage

\begin{center}
\begin{deluxetable}{ccccccl}
\tablecaption{The Total Cosmic X-ray Background Flux Density \label{tbl-cxrb_total}}
\tablewidth{0pt}
\tabletypesize{\normalsize}
\tablecolumns{7}\tablehead{
\colhead{}&
\colhead{Unresolved}&
\colhead{Unresolved}&
\colhead{Resolved}&
\colhead{Resolved}&
\colhead{Total}&
\colhead{}\\
\colhead{Band}&
\colhead{CXRB}&
\colhead{CXRB [\%]}&
\colhead{CXRB}&
\colhead{CXRB [\%]}&
\colhead{CXRB}&
\colhead{Reference}\\
\colhead{(1)}&
\colhead{(2)}&
\colhead{(3)}&
\colhead{(4)}&
\colhead{(5)}&
\colhead{(6)}&
\colhead{(7)}}
\startdata
0.5-2 & --- & --- & --- &   $  94.3^{+7.0  }_{-6.7  } $  & $  0.75^{+0.04 }_{-0.04 } $ & M03\\ 
 2-10 & --- & --- & --- &   $  88.8^{+7.8  }_{-6.6  } $  & $  2.02^{+0.11 }_{-0.11 } $ &    \\ 
\hline
0.5-2 & --- & --- & --- &   $  89.5^{+5.9  }_{-5.7  } $  & $  0.75^{+0.04 }_{-0.04 } $ & B04\\ 
  2-8 & --- & --- & --- &   $  92.6^{+6.6  }_{-6.3  } $  & $  1.79^{+0.11 }_{-0.11 } $ &    \\ 
\hline
1-2  & $  0.10^{+0.01 }_{-0.01 } $  & $  22.7^{+3.1  }_{-3.1  } $  & $  0.35^{+0.02 }_{-0.02 } $  & $  77.0^{+3.0  }_{-3.0  } $  & $  0.46^{+0.03 }_{-0.03 } $ & HM06 \\ 
2-8  & $  0.34^{+0.17 }_{-0.17 } $  & $  20.0^{+10.0 }_{-10.0 } $  & $  1.36^{+0.10 }_{-0.10 } $  & $  80.0^{+8.0  }_{-8.0  } $  & $  1.70^{+0.20 }_{-0.20 } $ &      \\ 
\hline
 0.5-2 & $  0.18^{+0.03}_{-0.03} $  & $  21.9^{+ 3.8}_{ -3.8} $  & $  0.63^{+0.01}_{-0.01} $  & $  78.1^{+ 1.2}_{ -1.2} $  & $  0.81^{+0.03}_{-0.03} $ & this study\\ 
 1-2 & $  0.10^{+0.01}_{-0.01} $  & $  21.5^{+ 2.9}_{ -2.9} $  & $  0.38^{+0.01}_{-0.01} $  & $  78.5^{+ 1.2}_{ -1.2} $  & $  0.48^{+0.02}_{-0.02} $ & without targets\\ 
 2-8 & $  0.34^{+0.17}_{-0.17} $  & $  19.5^{+ 9.8}_{ -9.8} $  & $  1.40^{+0.03}_{-0.03} $  & $  80.5^{+ 1.7}_{ -1.7} $  & $  1.74^{+0.17}_{-0.17} $ & \\ 
\hline
 0.5-2 & $  0.18^{+0.03}_{-0.03} $  & $  20.7^{+ 3.6}_{ -3.6} $  & $  0.68^{+0.01}_{-0.01} $  & $  79.3^{+ 1.2}_{ -1.2} $  & $  0.86^{+0.03}_{-0.03} $ & this study\\ 
 1-2 & $  0.10^{+0.01}_{-0.01} $  & $  20.2^{+ 2.7}_{ -2.7} $  & $  0.41^{+0.01}_{-0.01} $  & $  79.8^{+ 1.2}_{ -1.2} $  & $  0.51^{+0.02}_{-0.02} $ & with targets\\ 
 2-8 & $  0.34^{+0.17}_{-0.17} $  & $  18.5^{+ 9.2}_{ -9.2} $  & $  1.50^{+0.07}_{-0.07} $  & $  81.5^{+ 3.8}_{ -3.8} $  & $  1.84^{+0.18}_{-0.18} $ & \\ 
\enddata
\tablecomments{
Col. (1): X-ray energy band in keV. 
Col. (2): the unresolved CXRB flux density (HM05) 
in units of $10^{-11}$ $erg$ $cm^{-2}$ $sec^{-1}$ $deg^{-2}$.
Col. (3): the percentage of the total CXRB flux density that is unresolved.  
Col. (4): the resolved CXRB flux density from the ChaMP+CDFs number counts
in units of $10^{-11}~erg~cm^{-2}~sec^{-1}~deg^{-2}$. 
The resolved CXRB in the 1-2 keV band is rescaled from that 
in the 0.5-2 keV band assuming $\Gamma_{ph}=1.4$. 
Col. (5): the percentage of the total CXRB flux density that is resolved. 
Col. (6): the total CXRB flux density in units of 
$10^{-11}$ $erg$ $cm^{-2}$ $sec^{-1}$ $deg^{-2}$.
This column is the sum of the column (2) and the column (4) for HM05 and this study.
Col. (7): reference
}
\end{deluxetable}
\end{center}
\clearpage

\plotone{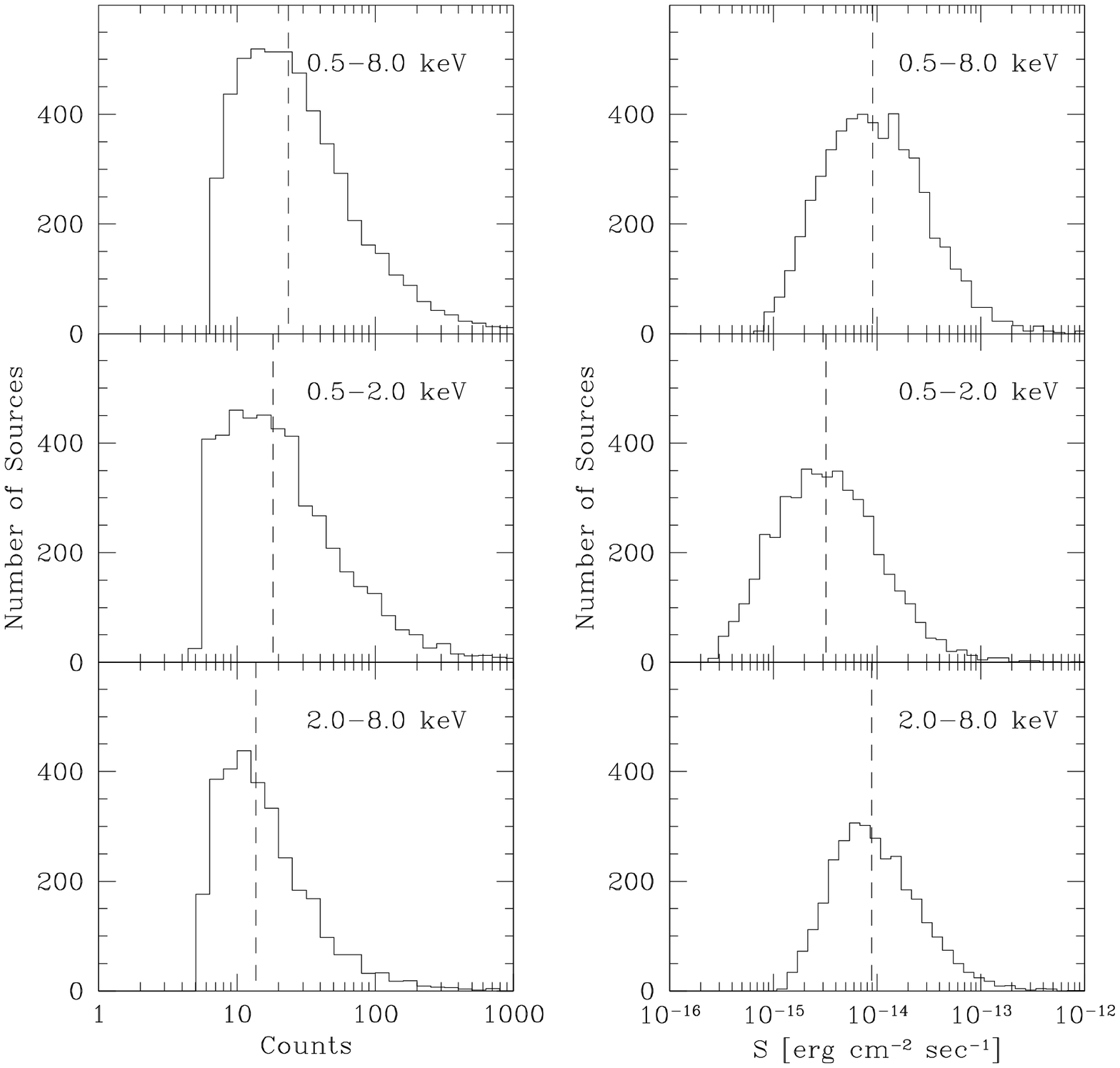}
\figcaption[f1.eps]{
The distributions of source net counts ($left$) and flux ($right$) 
in three energy bands.
The vertical dashed line indicates the median of the 
distribution.
The flux was determined assuming a photon index of $\Gamma_{ph}=1.4$.
\label{fig-count_flux}}

\plotone{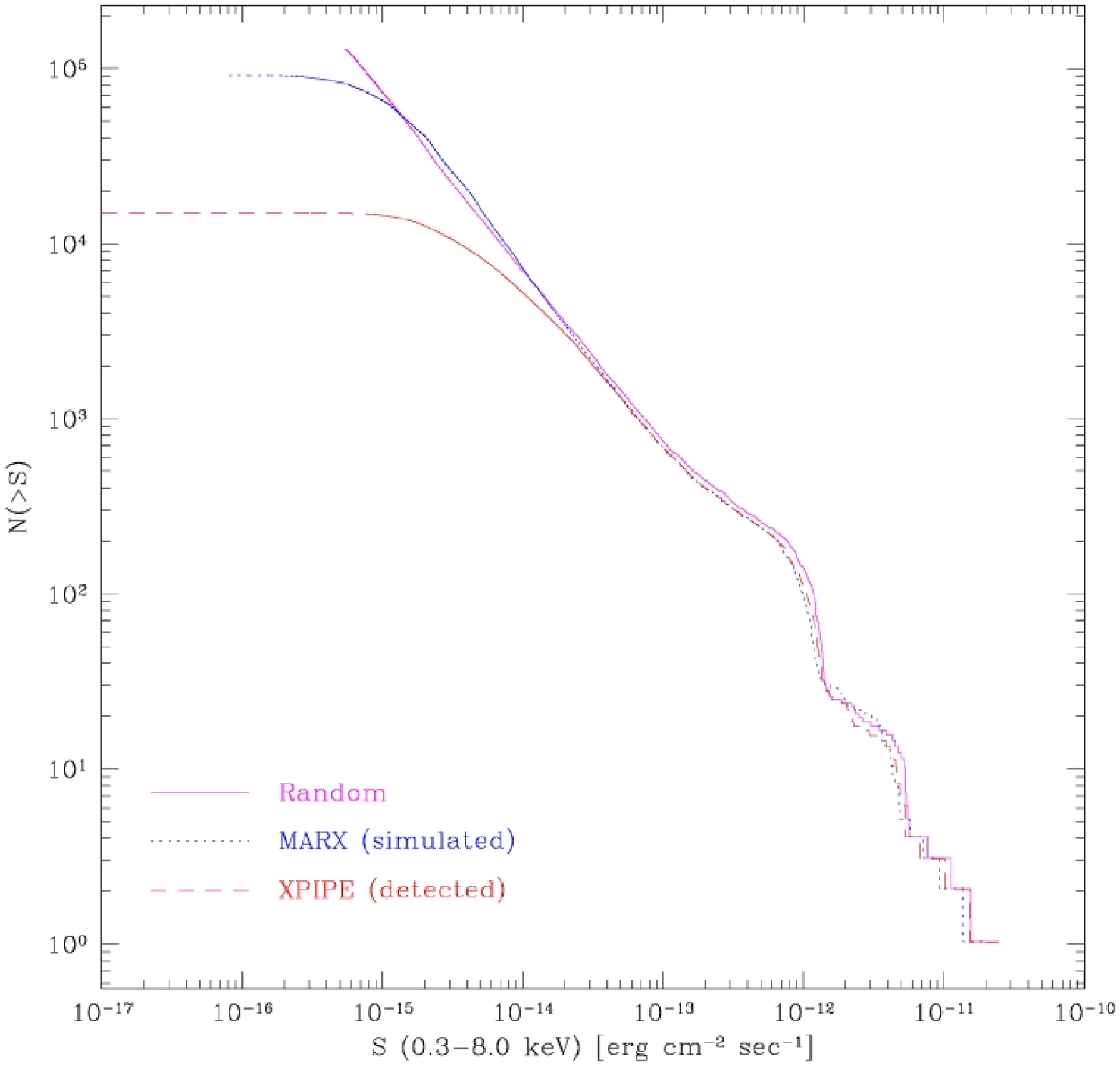}
\figcaption[f2.eps]{
The cumulative number counts for the artificial sources in 
the B band. 
The magenta solid line represents the number counts for 
sources whose fluxes were randomly selected from the assumed
number counts with a slope of $-1$.
Due to small number statistics, deviations from the assumed number counts
are present in the bright flux regime.
Blue dotted and red dashed lines represent number counts for 
sources generated with MARX 
and for sources extracted with XPIPE, respectively. 
The effect of Eddington bias is evident at the faint fluxes 
($S$ $<$ $10^{-14}$ $erg$ $cm^{-2}$ $sec^{-1}$) in the simulated
source number counts.
\label{fig-cflux}}

\plotone{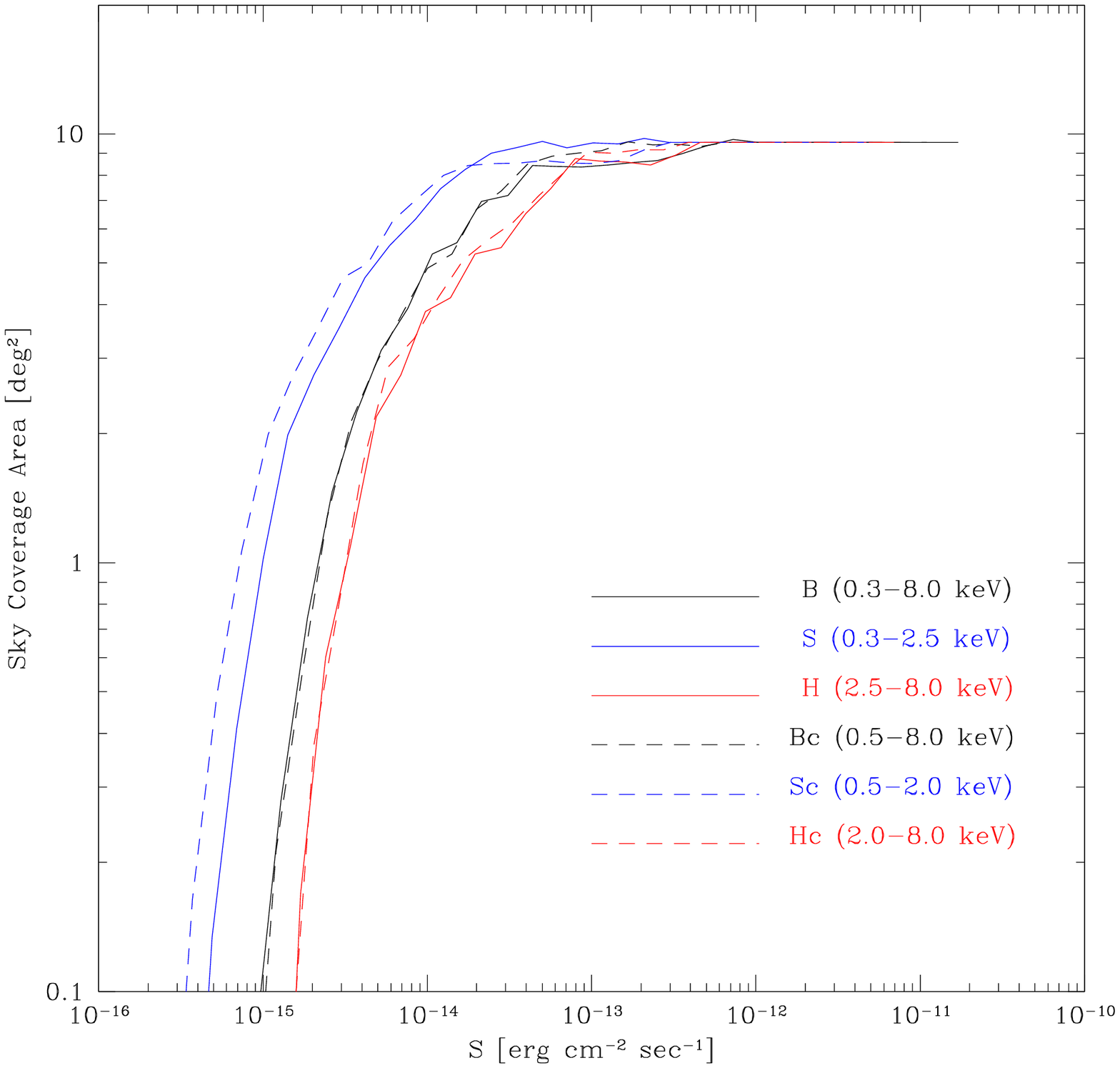}
\figcaption[f3.eps]{
Sky coverages for sources with $S/N>1.5$ in six energy bands. 
The full sky coverage is 9.6 $deg^{2}$.
\label{fig-sc}}

\plotone{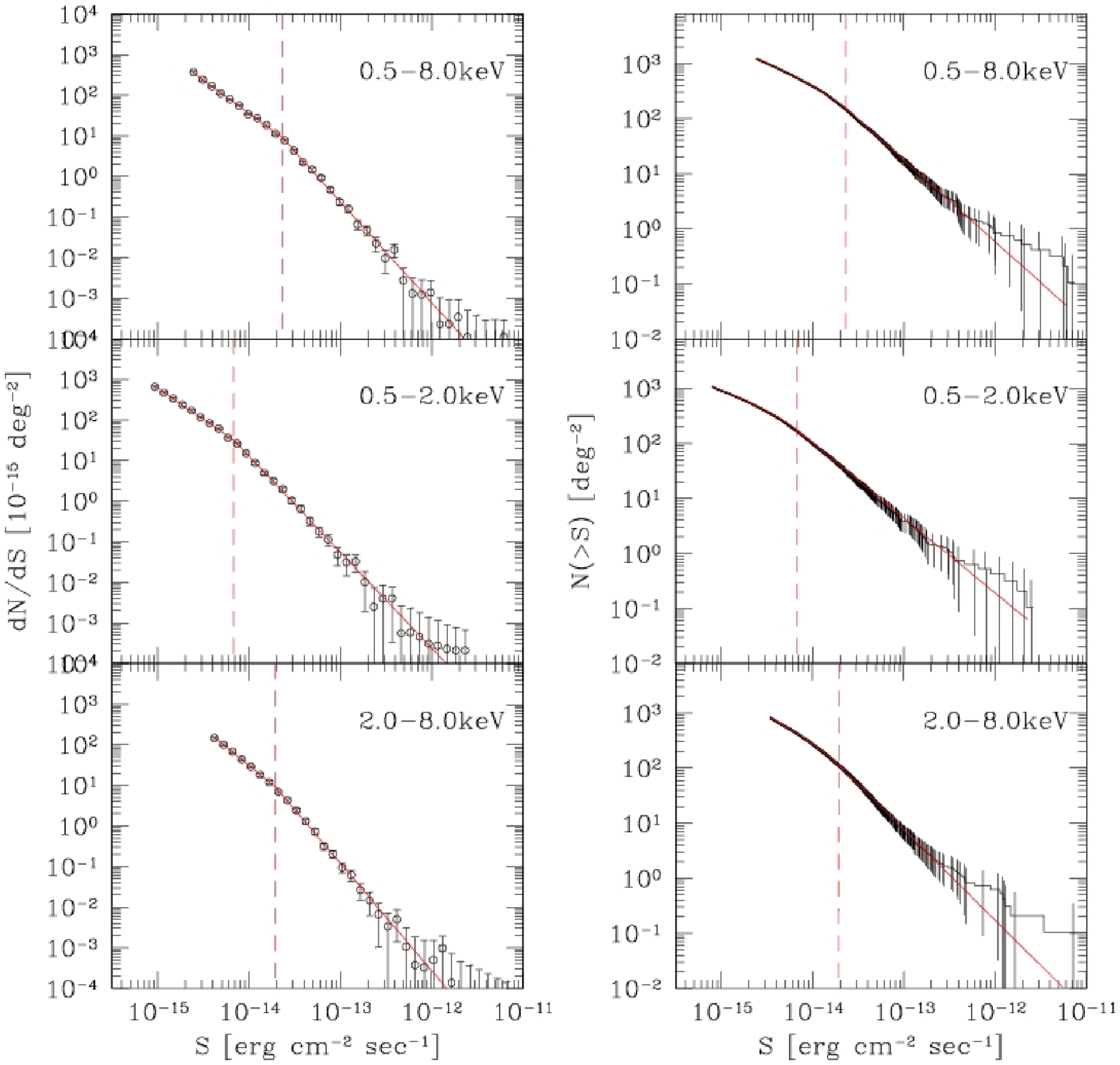}
\figcaption[f4.eps]{
The differential ($left$) and cumulative ($right$) number counts
of the ChaMP X-ray point sources in the Bc, Sc, and Hc bands, respectively.
Red solid lines represent the best fit results with a broken power law.
The vertical red dashed lines indicate the derived break fluxes.
Source fluxes were determined assuming a photon index of $\Gamma_{ph}=1.4$.
Since we present the differential number counts brighter than a flux 
corresponding to $10\%$ of the full sky coverage,
the faintest bin still has sufficient sources and shows a 
small error bar. 
The error bars in the cumulative number counts are estimated by  
the error propagation rule using \citet{geh86} statistics. 
\label{fig-sherpa_fit_C}}

\plotone{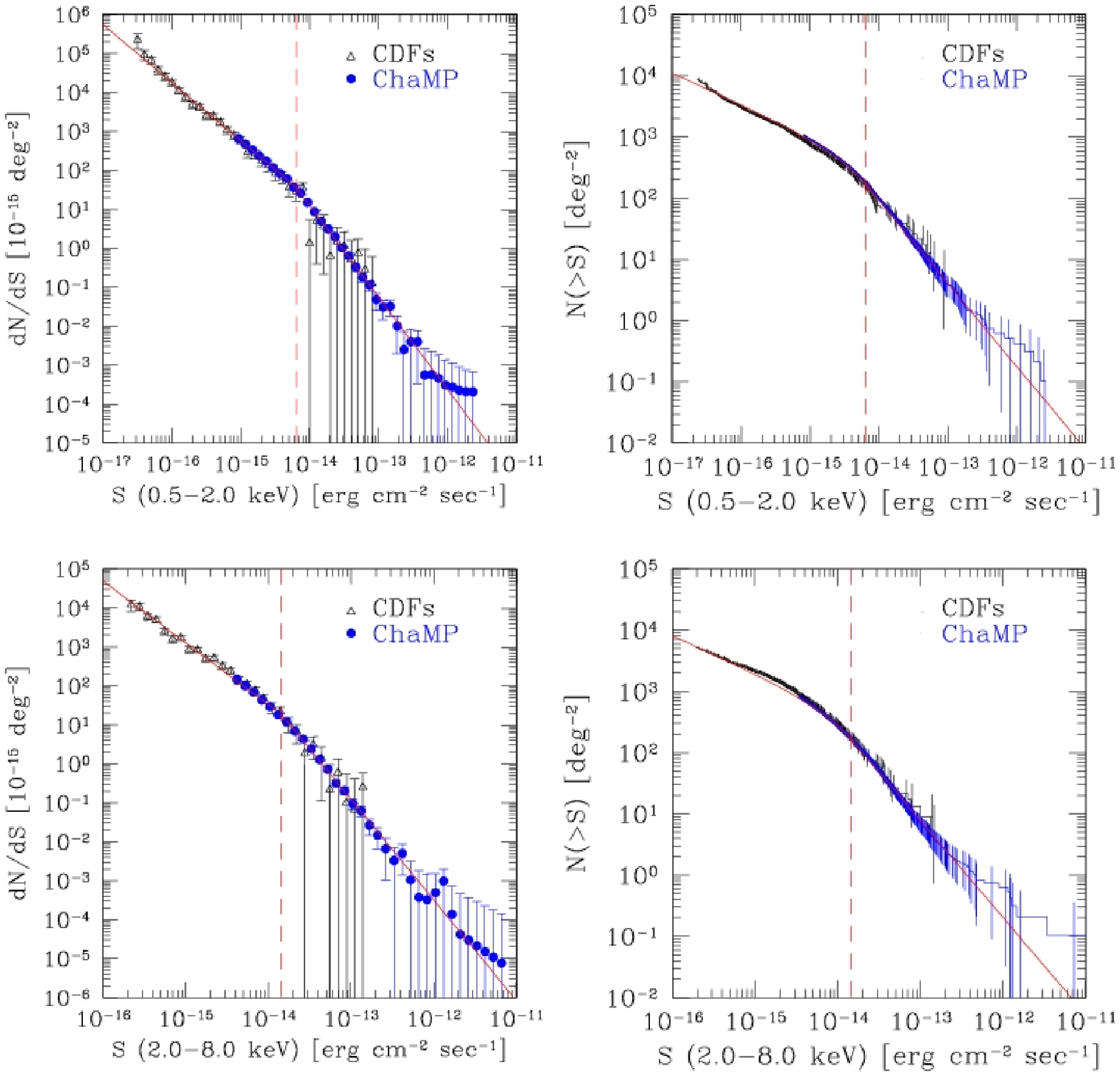}
\figcaption[f5.eps]{
The differential ($left$) and cumulative ($right$) number counts 
for the ChaMP+CDFs X-ray point sources in the Sc and Hc bands. 
Blue filled circles and black open triangles represent 
the ChaMP and the CDFs data,
respectively.
Red lines are the best simultaneous fit results.
The vertical dashed lines indicate the derived break fluxes.
Source fluxes were determined assuming a photon index of $\Gamma_{ph}=1.4$.
\label{fig-ncount_fit_wcdf}}

\plotone{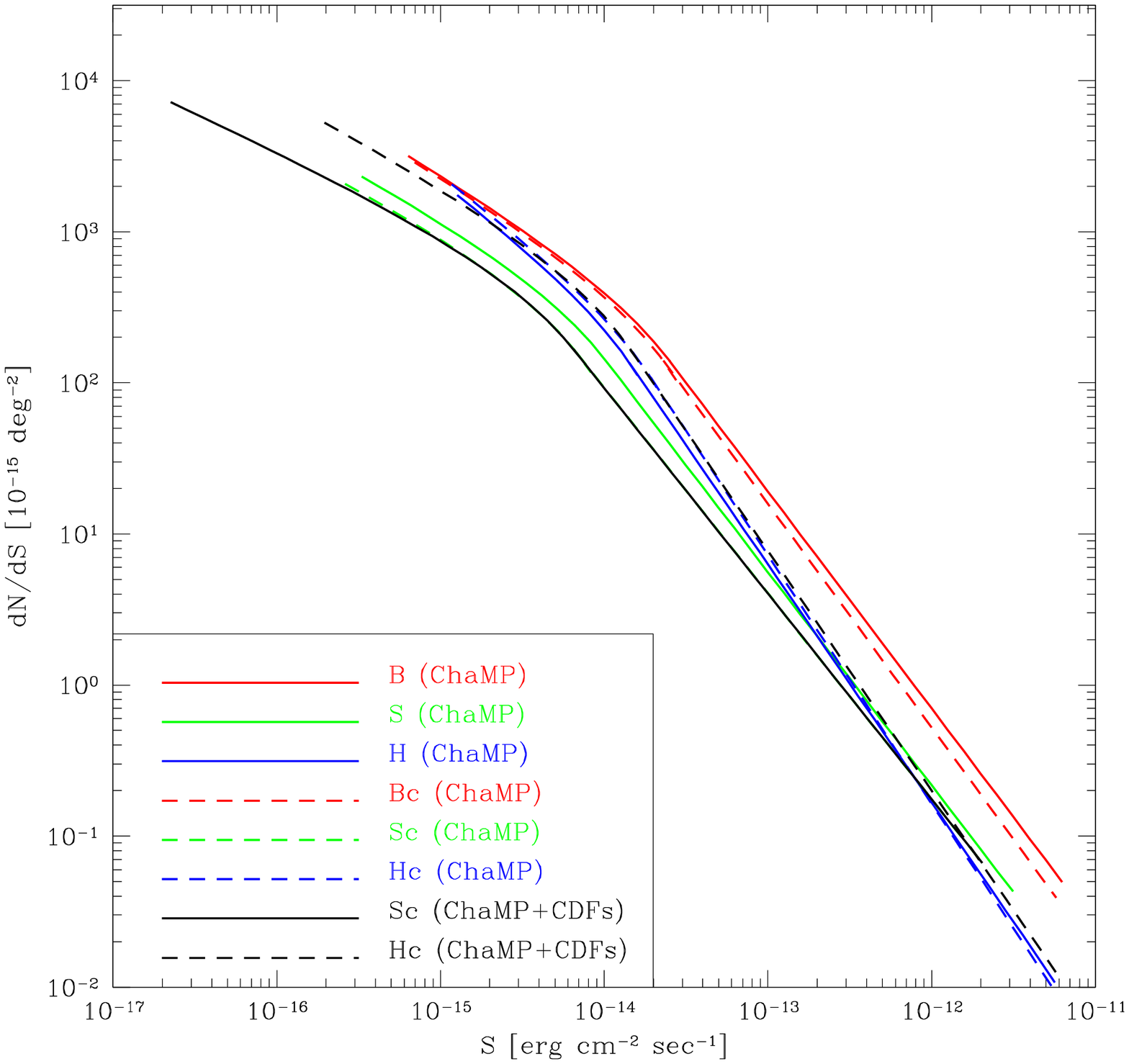}
\figcaption[f6.eps]{
The differential number counts of the ChaMP and the ChaMP+CDFs 
from the best fit results in 6 energy bands.
Source fluxes were determined assuming a photon index of $\Gamma_{ph}=1.4$. 
For energy band definition, see Table \ref{tbl-energy-def}.
\label{fig-sherpa_fit_6}}

\plotone{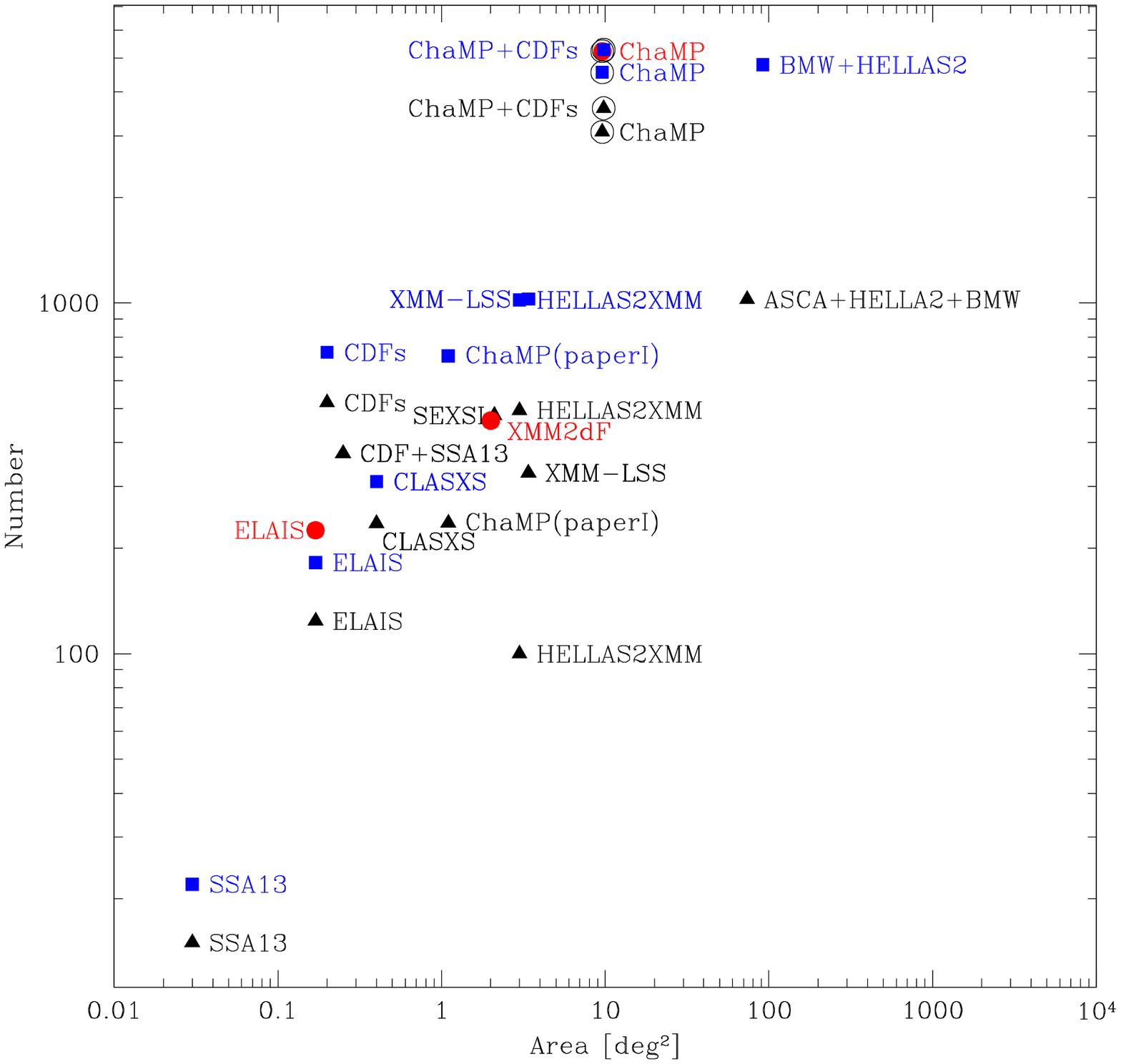}
\figcaption[f7.eps]{
The number of sources and covered sky areas for various studies.
Red circles, blue squares, and black triangles represent
the broad, soft, and hard energy bands, respectively.
References and parameters are listed in Table \ref{tbl-other}. 
The ChaMP contains $\sim5,200$ sources in the 0.5-8 keV band
and covers $9.6$ $deg^{2}$ in sky area, and the ChaMP+CDFs 
covers $9.8$ $deg^{2}$.
For this study, the 0.5-2 keV, 2-8 keV, and 0.5-8 keV bands 
are used and are marked by open circles for clear comparison.
\label{fig-area_other}}

\plotone{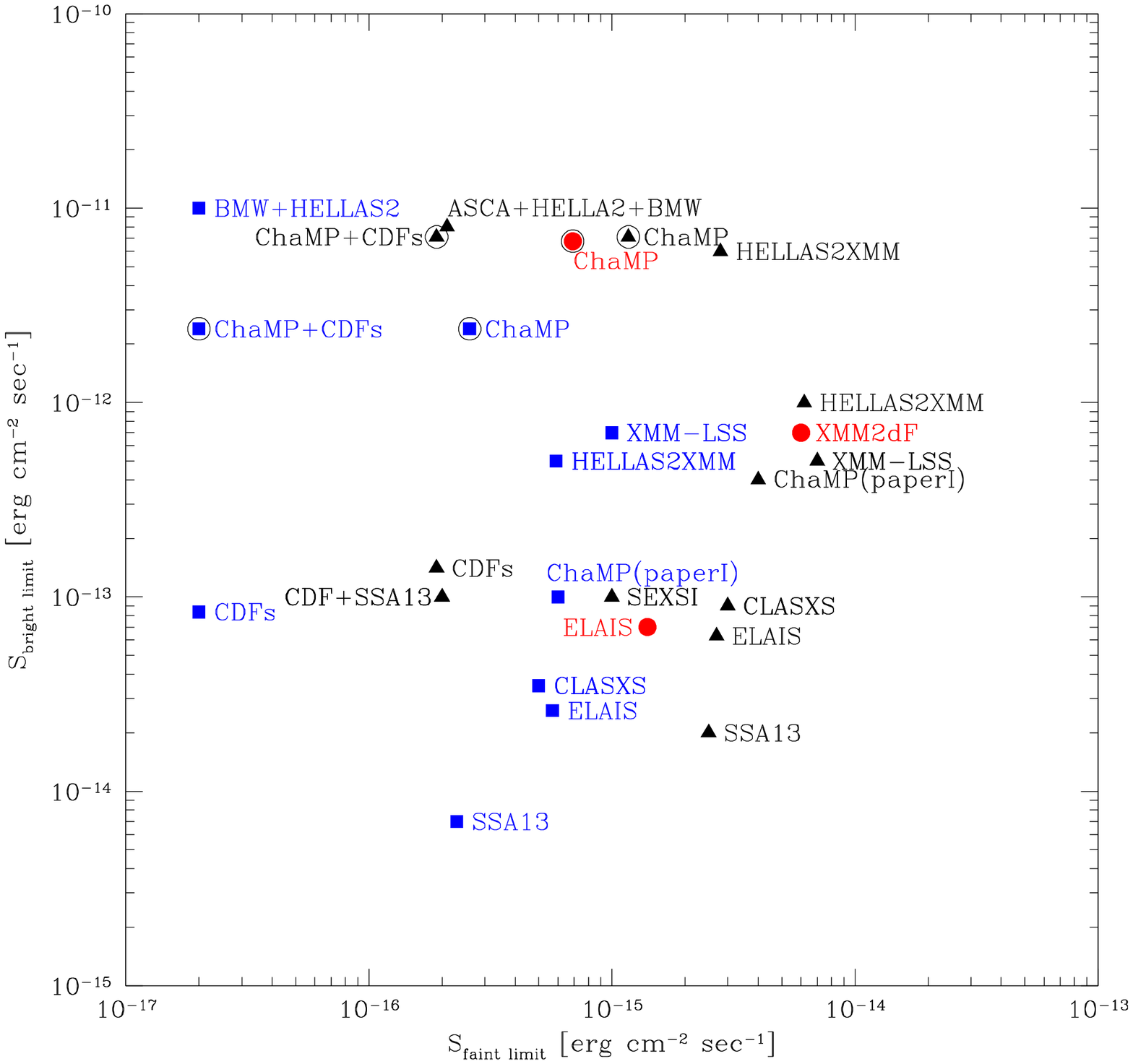}
\figcaption[f8.eps]{
The faint and bright flux limits of various studies.
Red circles, blue squares, and black triangles represent
the broad, soft, and hard energy bands.
References and parameters are listed in Table \ref{tbl-other}.
For this study, the 0.5-2 keV, 2-8 keV, and 0.5-8 keV bands 
are used and are marked by open circles for clear comparison.
We note that the faint and bright flux limit of the ChaMP are 
estimated from ChaMP sources with $S/N>1.5$.
For the other studies, the faint and bright flux limits are from 
the literature or from our own visual estimations based on their    
number counts.
\label{fig-flux_other}}

\plotone{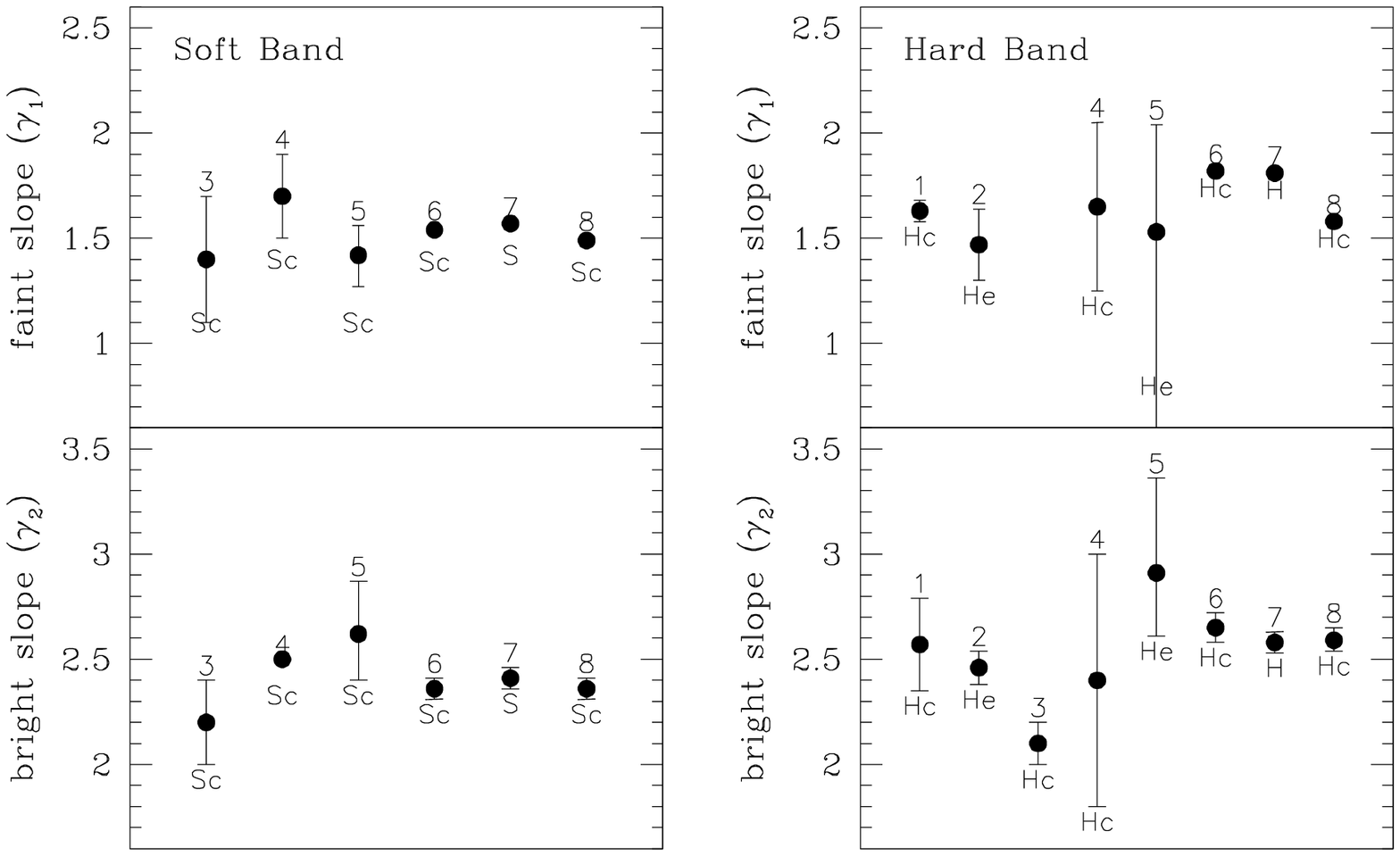}
\figcaption[f9.eps]{
The faint ($top$) and bright ($bottom$) power indices of the
differential number counts for this and 
previous studies in the soft ($left$) and the hard ($right$) 
bands, respectively. The references are marked by number: 1: \citet{cow02},
2: H03, 3: KD04, 4: \citet{yan04},
5: \citet{chi05}, 6-7: This study for the ChaMP, and 8: This study for the ChaMP+CDF.
The energy bands of each study are also marked here
and for the definition of them, see Table \ref{tbl-energy-def}.
Note that \citet{yan04} fixed the bright slope as 2.5 in the Sc band 
having no error.  
\label{fig-gamma_other}}

\plotone{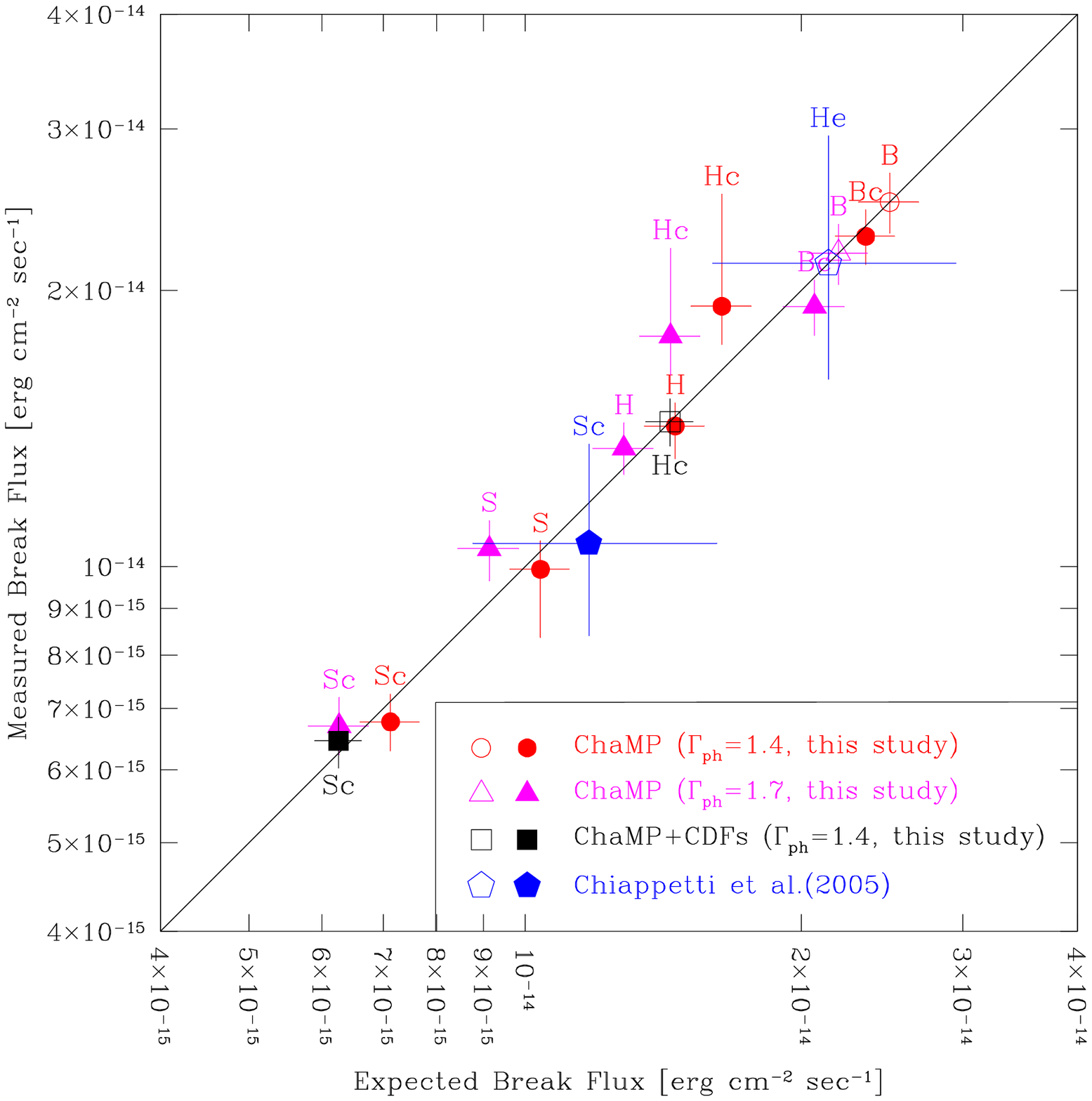}
\vspace{-15mm}
\figcaption[f10.eps]{
Comparison of the expected and the measured break fluxes of
the differential number counts in various energy bands.
Assuming a single power law model
for the X-ray spectrum, 
the expected break flux in each band ($closed$ $symbol$) was calculated by converting 
the $S_{b},std$ (see equation (7))
which is the measured break flux in the B (ChaMP, $open$ $circle$ and $triangle$), 
Hc (ChaMP+CDFs, $open$ $square$),
or He (\citet{chi05}, $open$ $pentagon$) band. 
For the ChaMP, photon indices of 
$\Gamma_{ph}=1.4$ ($circles$) and $\Gamma_{ph}=1.7$ ($triangles$) were assumed.
A photon index of $\Gamma_{ph}=1.7$ is assumed for \citet{chi05} 
for self-consistency with their study.
The solid line represents the line of equality for expected and measured
break fluxes and is shown for illustrative comparison.
The energy bands are listed in Table \ref{tbl-energy-def}.
\label{fig-bflux}}

\plotone{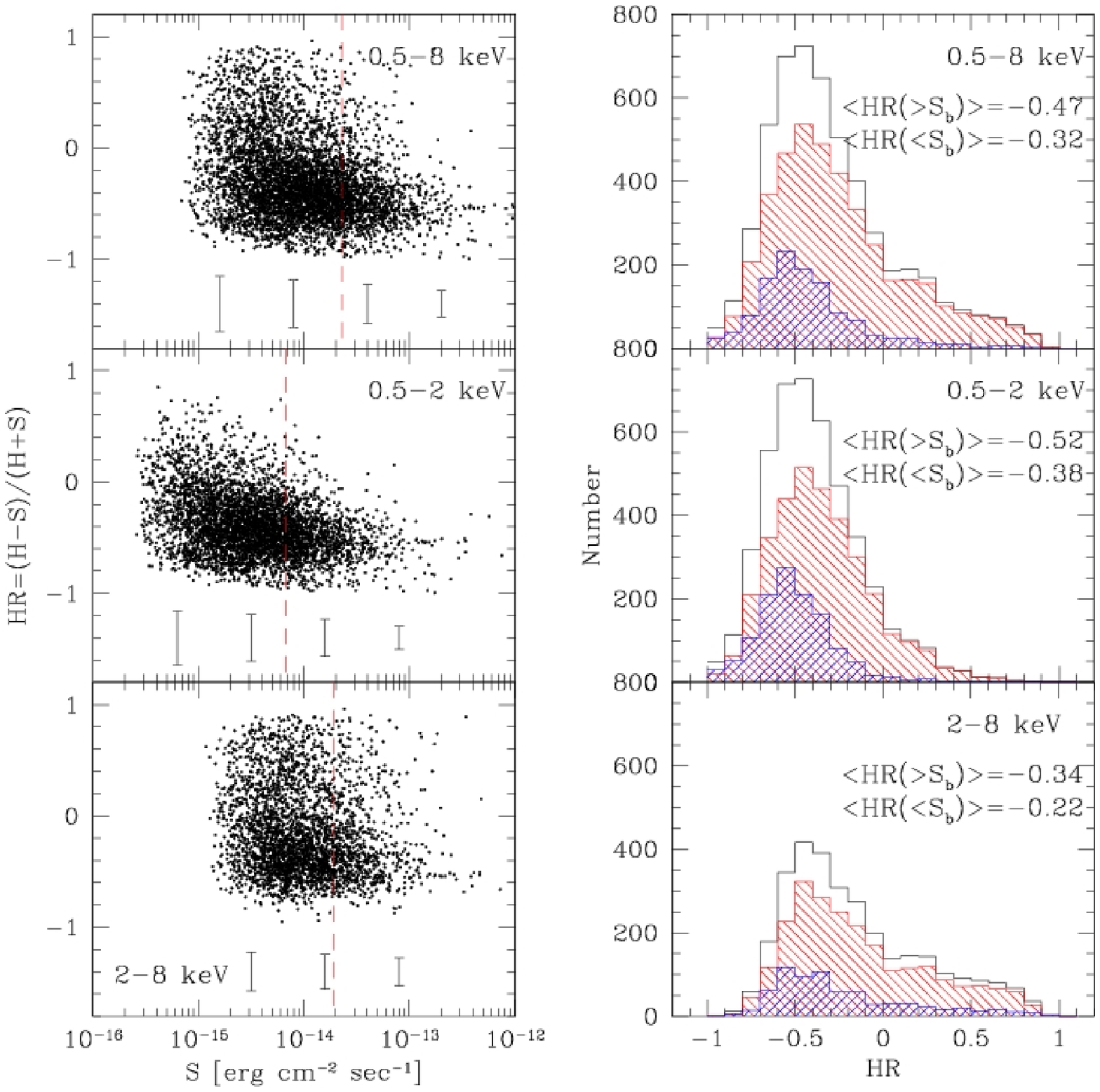}
\vspace{-10mm}
\figcaption[f11.eps]{
$left.$ Hardness ratio (HR) distributions as a function of flux
in three energy bands. 
A photon index of $\Gamma_{ph}=1.4$ was assumed. 
Error bars plotted at the bottom of each panel are the 
typical uncertainties of the hardness
ratio at several flux bins. 
The vertical red dashed line indicates the break flux 
($S_{b}$ in Table \ref{tbl-fit_diff_nt})
in each energy band.
$right.$ Hardness ratio distributions in the following 
flux ranges: entire flux range ($open$ $histogram$),
$S<S_{b}$ ($red$ $shaded$ $histogram$),
and $S>S_{b}$ ($blue$ $shaded$ $histogram$). 
The median hardness ratio of the faint and bright samples
are marked in each panel.  
\label{fig-flux_hr}}

\plotone{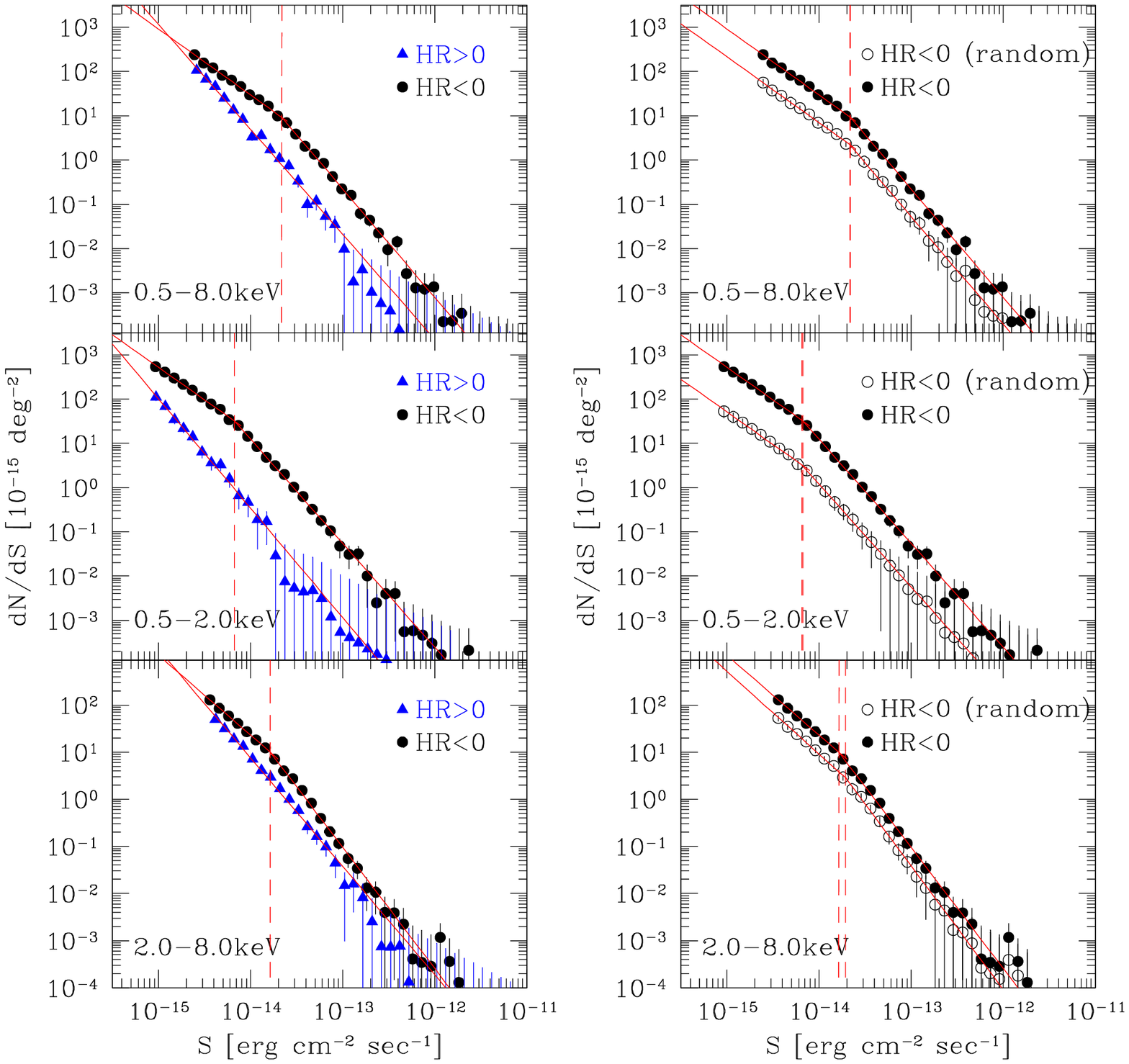}
\vspace{-15mm}
\figcaption[f12.eps]{
$left.$ Differential number counts for the soft 
(HR$<0$, $black$ $circles$) and hard (HR$>0$, $blue$ $triangles$)
sources in 3 energy bands.
Soft sources show a
break and are fitted with a broken power law 
while hard sources
do not show a break and are fitted with a single power law.
Red solid lines are the best fit results 
(see Table \ref{tbl-fit_diff_nt_hr}).
The red vertical dashed line is the break flux 
in the soft source number counts. 
$right.$ Averaged differential number counts from $1,000$
random subsets of soft sources ($open$ $circles$). 
Each subset has the same number of soft sources as hard sources. 
Even with the reduced statistics, soft sources still show 
a significant break and are fitted with a broken power law. 
The number counts for all soft sources ($filled$ $circles$) 
is plotted in each energy band for comparison.
\label{fig-sub_hr_ran}}

\plotone{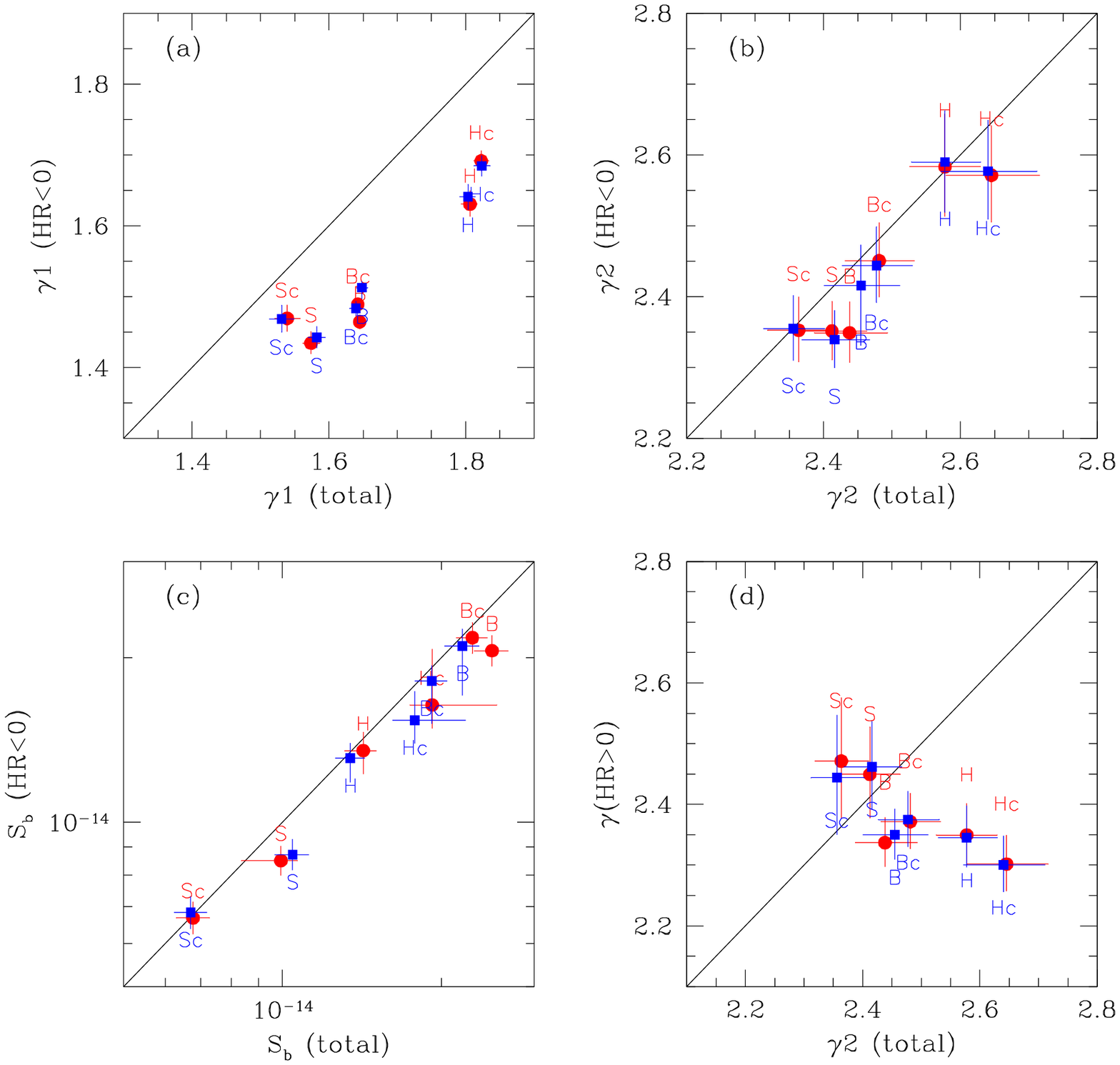}
\vspace{-15mm}
\figcaption[f13.eps]{
Comparison of the best fit parameters of the ChaMP differential
number counts for the total sample
(see Table \ref{tbl-fit_diff_nt})
with those for subsamples (see Table \ref{tbl-fit_diff_nt_hr}).
The total sample includes all sources regardless of the HR,
the soft sample includes sources with HR$<0$, and the
hard sample includes sources with HR$>0$.
(a) Faint power law indices of
the total sample vs. the soft sample.
(b) Bright power law indices of the total sample vs. the soft sample.
(c) Break flux of the total sample vs. the soft sample.
(d) Bright power law indices of the total sample vs. 
single power law indices of 
the hard sample.
The photon indices of $\Gamma_{ph}=1.4$ ($red$ $circles$)
and $\Gamma_{ph}=1.7$ ($blue$ $squares$) are assumed.
The solid line represents the line of equality for the two compared parameters 
and is shown for illustrative comparison.
\label{fig-comp_index}}

\plotone{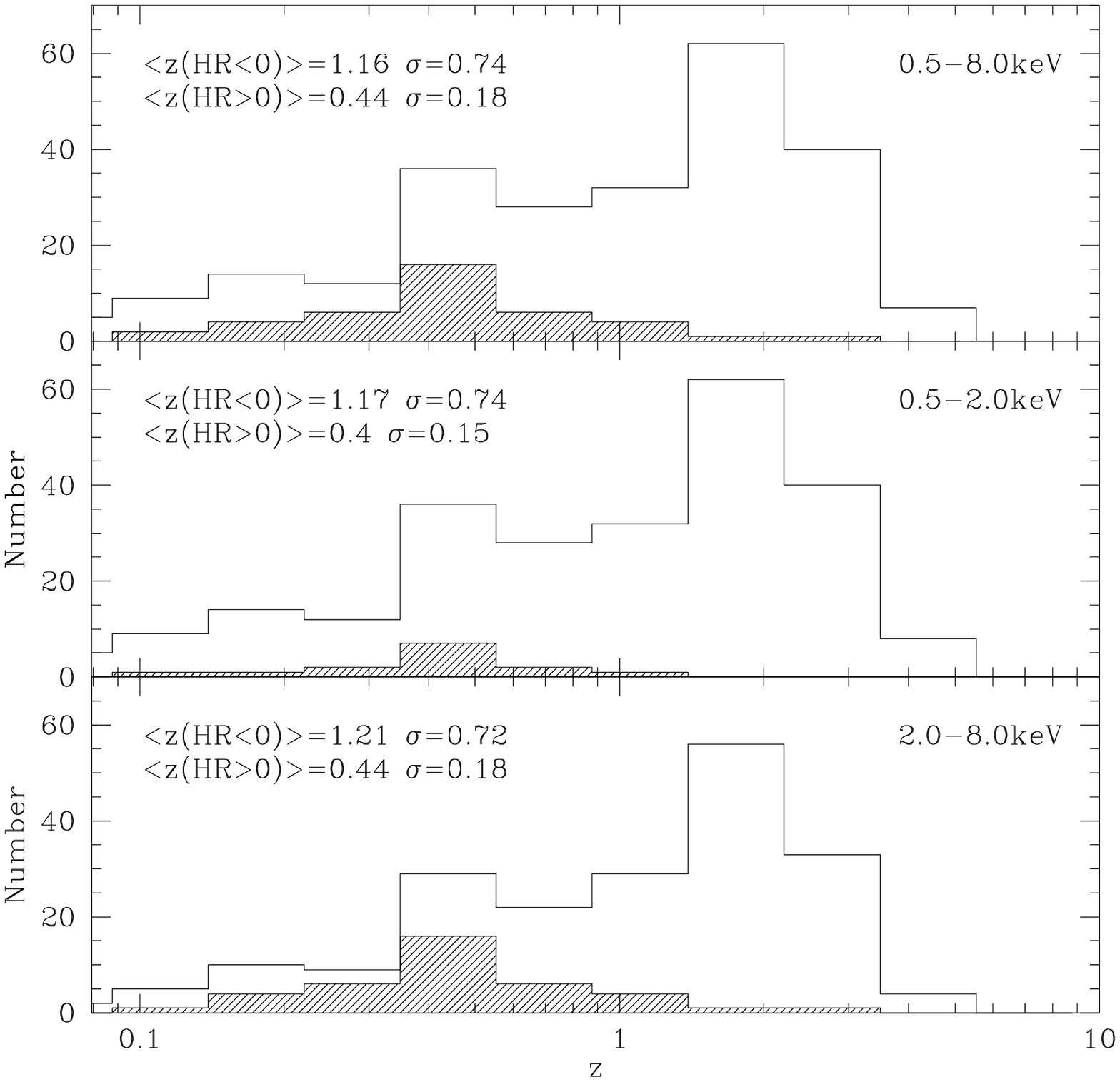}
\figcaption[f14.eps]{
The redshift distributions of the 
soft (HR$<0$, $open$ $histogram$) 
and hard (HR$>0$, $shaded$ $histogram$) sources 
in the Bc ($top$), Sc ($middle$), and Hc ($bottom$) bands, 
respectively. 
The medians and standard deviations for each distribution
are marked. 
Hard sources are distributed at lower redshifts compared
to the soft sources in all energy bands. 
Sources with the highest optical counterpart match confidence 
levels and with 
the highest spectrum quality are only used. 
\label{fig-redshift_hr1}}

\plotone{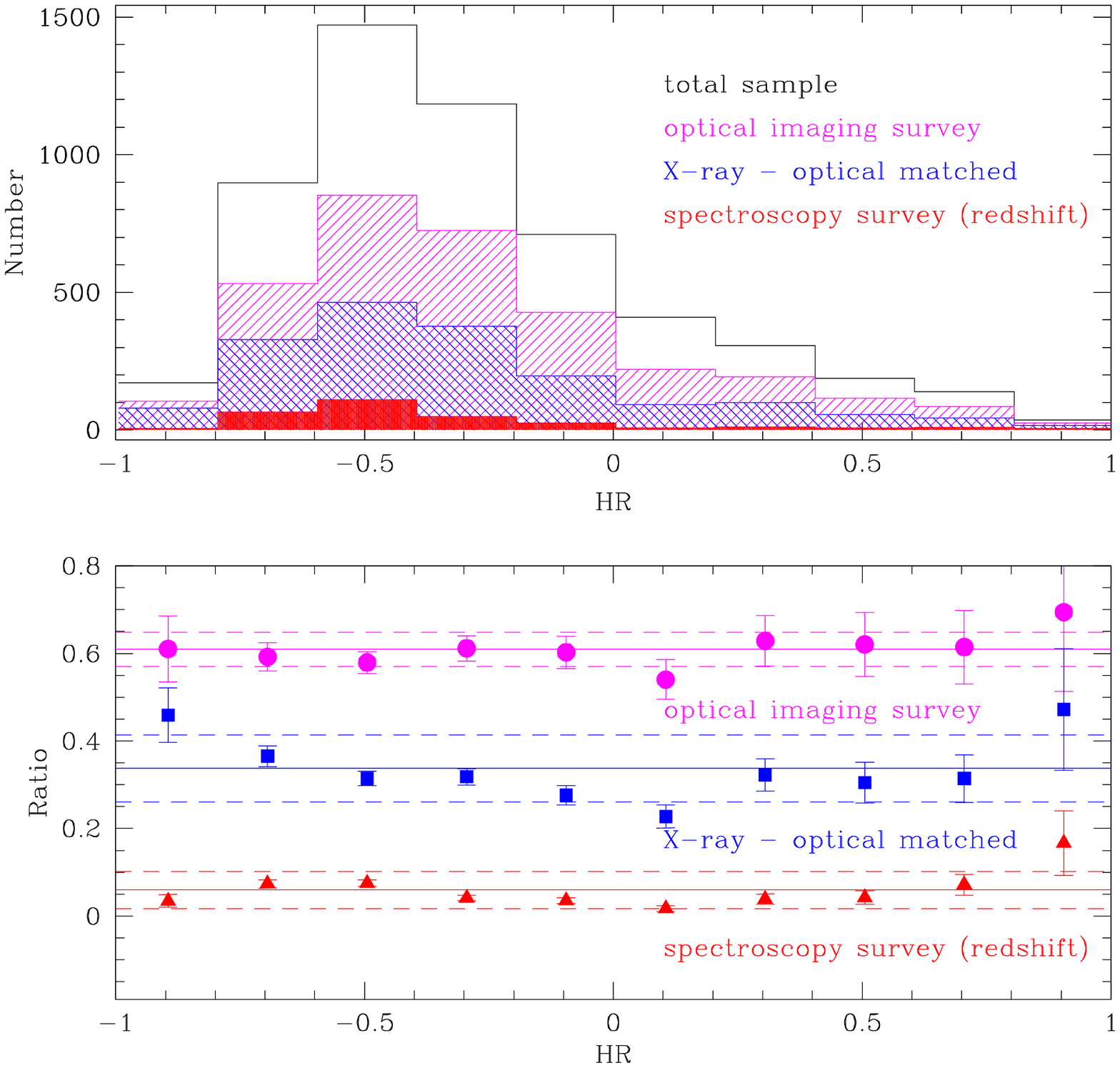}
\vspace{-10mm}
\figcaption[f15.eps]{
$top.$ Hardness ratio distribution of the ChaMP sources
in the B (0.3-8 keV) band in the following categories:
total sample ($open$ $histogram$),
sources in 63 ChaMP
fields with optical imaging observations ($magenta$ $shade$ $histogram$),
sources having optical
counterparts ($blue$ $shade$ $histogram$),
and sources having redshifts ($red$ $filled$ $histogram$).
$bottom.$ The number ratios of the last three subsamples
over total sample in each hardness
ratio bin. The mean ($solid$ $line$) and standard deviations
($dashed$ $lines$) of each ratio are plotted.
For the X-ray-optical matched sample and the redshift sample, only sources
with the highest match confidence levels and with
the highest spectrum quality are used.
\label{fig-imat_hr}}

\plotone{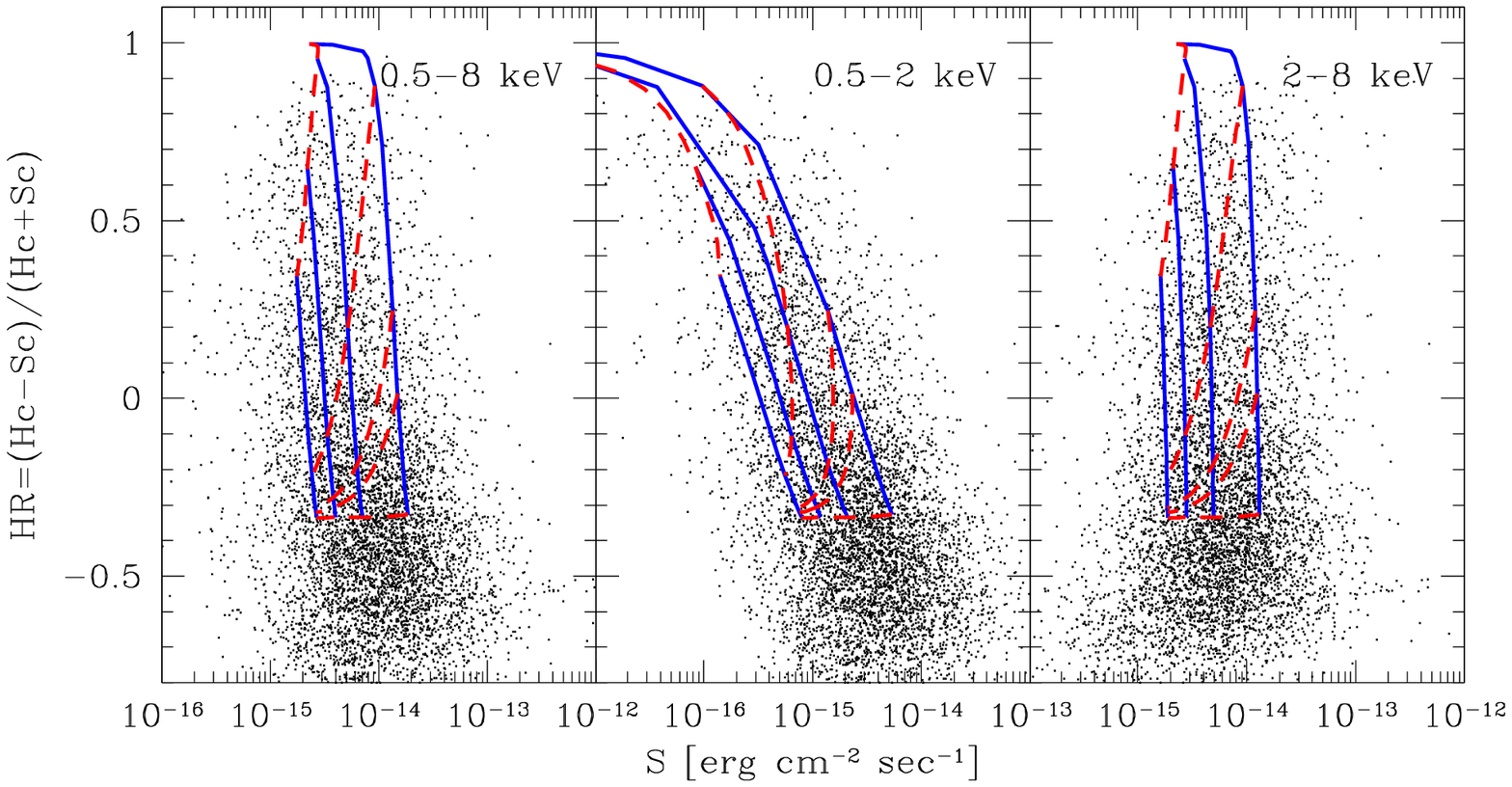}
\vspace{-20mm} 
\figcaption[f16.eps]{
The flux-hardness ratio (S-HR) diagram for the ChaMP 
sources with $S/N>0$ ($balck$ $dots$) in the 0.5-8 keV ($left$), 
0.5-2 keV ($middle$),
and 2-8 keV ($right$), respectively.
The grid indicates the predicted location of a 
test X-ray source for various redshift
($z=$0, 1, 2, and 3, blue solid lines from right to left)
and intrinsic absorption
($logN_{H,int}=$20, 21.7, 22, 22.7, and 23.7, 
red dashed lines from bottom to top).
A photon index of $\Gamma_{ph}=1.4$ was assumed
for the test source spectrum.
The source becomes fainter with increasing intrinsic absorption
and with increasing redshift.
The source becomes harder with increasing intrinsic absorption;
but, softer with increasing redshift.
These effects are significant in the soft band.
We note that the grid illustrates only  
a test source to indicate trends but does not cover the full range of
the ChaMP sources. 
\label{fig-hr_flux}}

\plotone{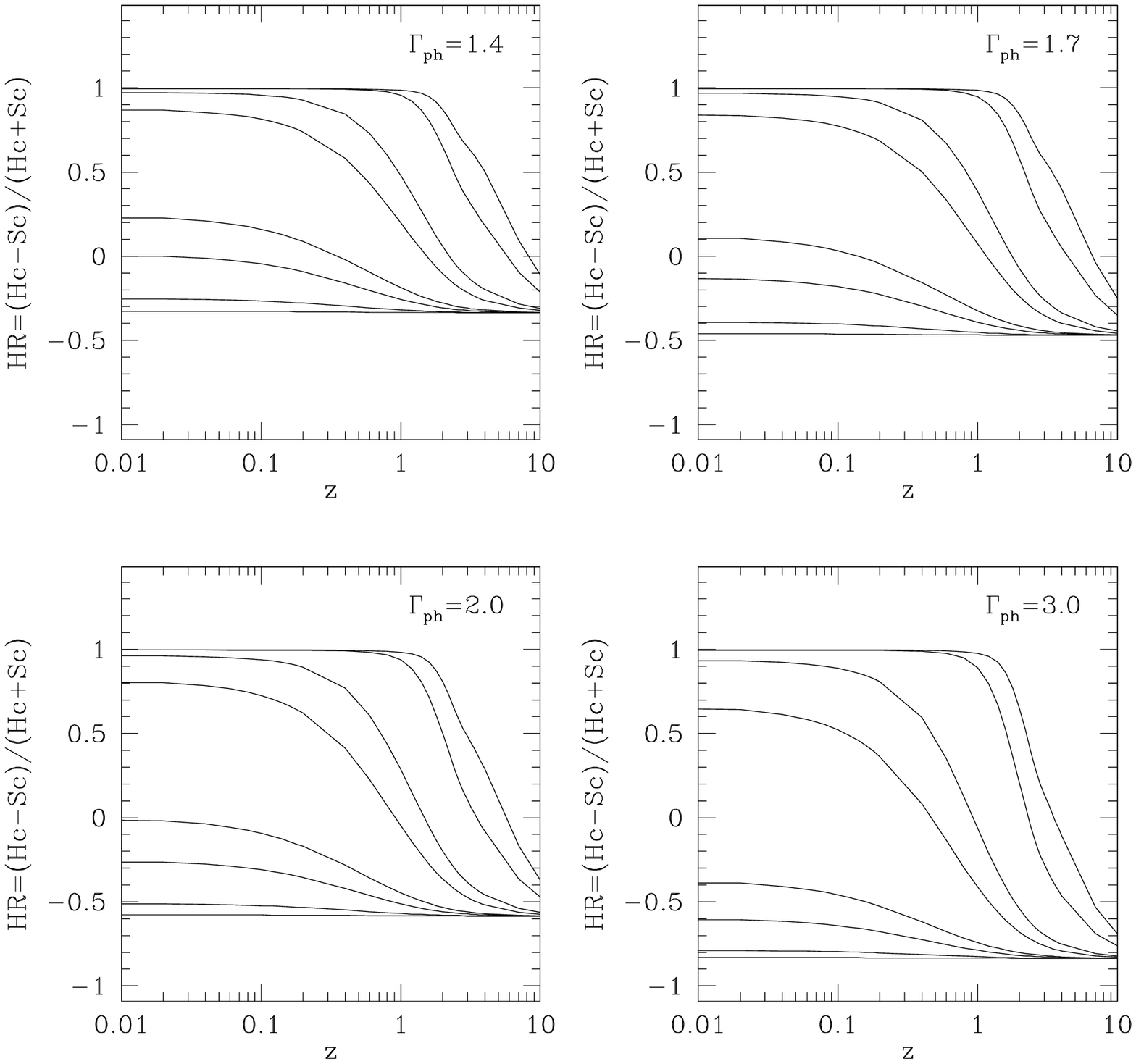}
\figcaption[f17.eps]{
The hardness ratio of the test X-ray source as a function of redshift for
photon indices of $\Gamma_{ph}=1.4$ ($top$ $left$), 
1.7 ($top$ $right$), 2.0 ($bottom$ $left$), 
and 3.0 ($bottom$ $right$), respectively.
In each panel, seven lines represent intrinsic absorptions
of $logN_{H,int}=$20, 21, 21.7, 22, 22.7, 23, 23.7, and 24, respectively,
from bottom to top.
The test source becomes harder with increasing intrinsic absorption
and with increasing redshift.
The intrinsically hard source with high redshift
is observed as a soft source in the observed frame due to 
the cosmological redshift.
\label{fig-hr_z}}

\plotone{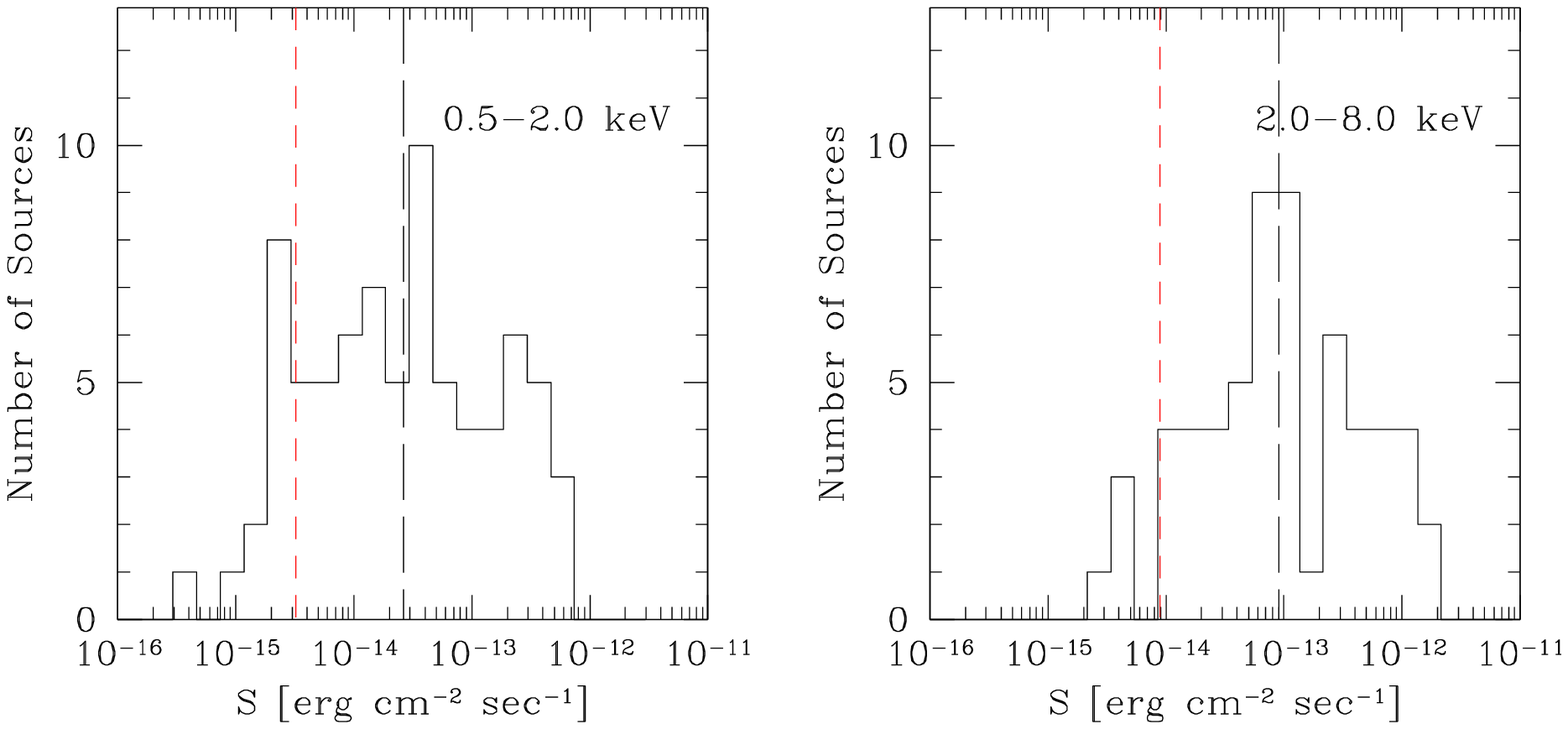}
\figcaption[f18.eps]{
The flux distribution of target sources in the Sc ($left$)
and the Hc ($right$) bands, respectively.
The vertical lines indicate the median values of 
target ($black$ $long$ $dashed$ $line$) and non-target
($red$ $short$ $dashed$ $line$) 
sources (see the left panels of 
Figure \ref{fig-count_flux} and Table \ref{tbl-prop}
for non-target sources), respectively.
Target sources have brighter fluxes compared with non-target sources.
\label{fig-flux_target}}

\plotone{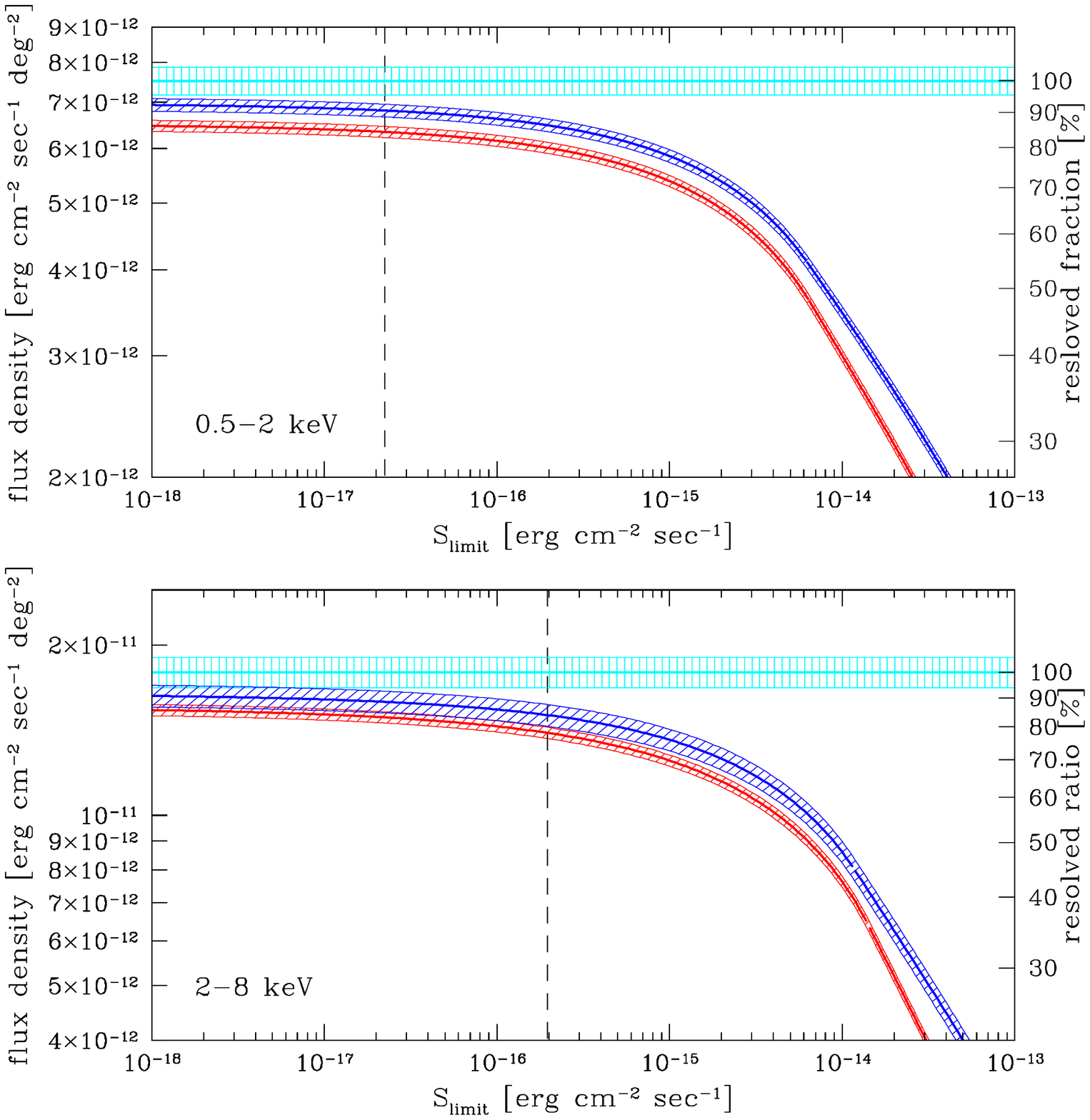}
\figcaption[f19.eps]{
The resolved CXRB flux density from the ChaMP+CDFs number counts
as a function of flux limit in the 0.5-2 keV ($top$)
and 2-8 keV ($bottom$) bands.
Blue and red shading represents the resolved CXRB with and 
without targets within $\pm1\sigma$
uncertainties, respectively.
The cyan shaded level represents the total CXRB and the $\pm1\sigma$
confidence range from the literature (B04). 
The vertical dashed line indicates
the faint flux limit of the ChaMP+CDFs. Below the faint flux limits, 
the results are extrapolated.
A photon index of $\Gamma_{ph}=1.4$ was assumed.
\label{fig-cxrb_SC_HC}}

\plotone{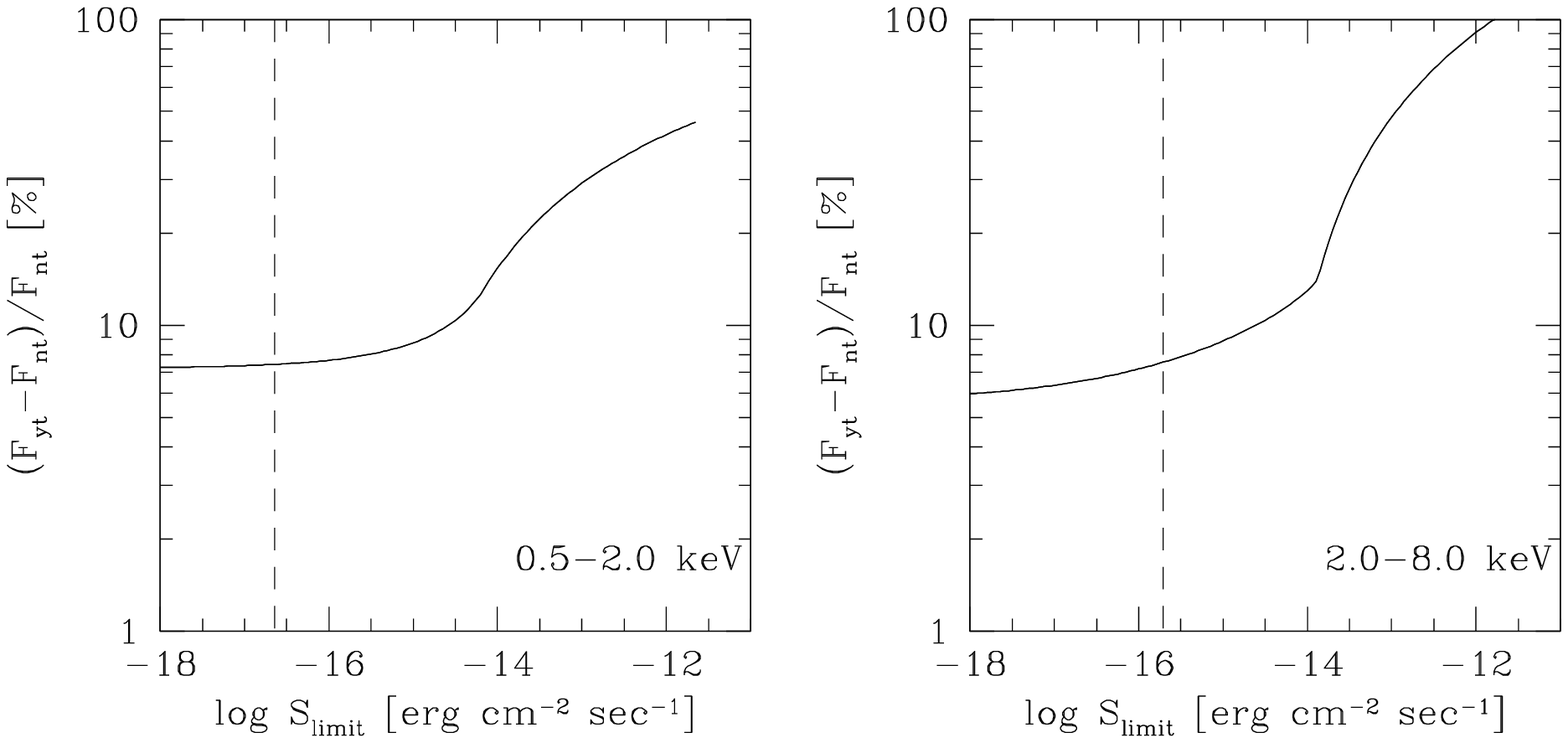}
\figcaption[f20.eps]{
The fractional difference between the resolved
CXRB excluding ($F_{nt}$) and including ($F_{yt}$) target sources, 
normalized to that of excluding targets
in the 0.5-2 keV ($left$)
and in the 2-8 keV ($right$) bands, respectively.
The differences are extrapolated below the faint flux limits
(vertical dashed lines).
At the faint flux limits, the target source fractions are
$7\%$ (0.5-2 keV) and $6\%$ (2-8 keV), respectively.
\label{fig-cxrb_6_comp}}

\plotone{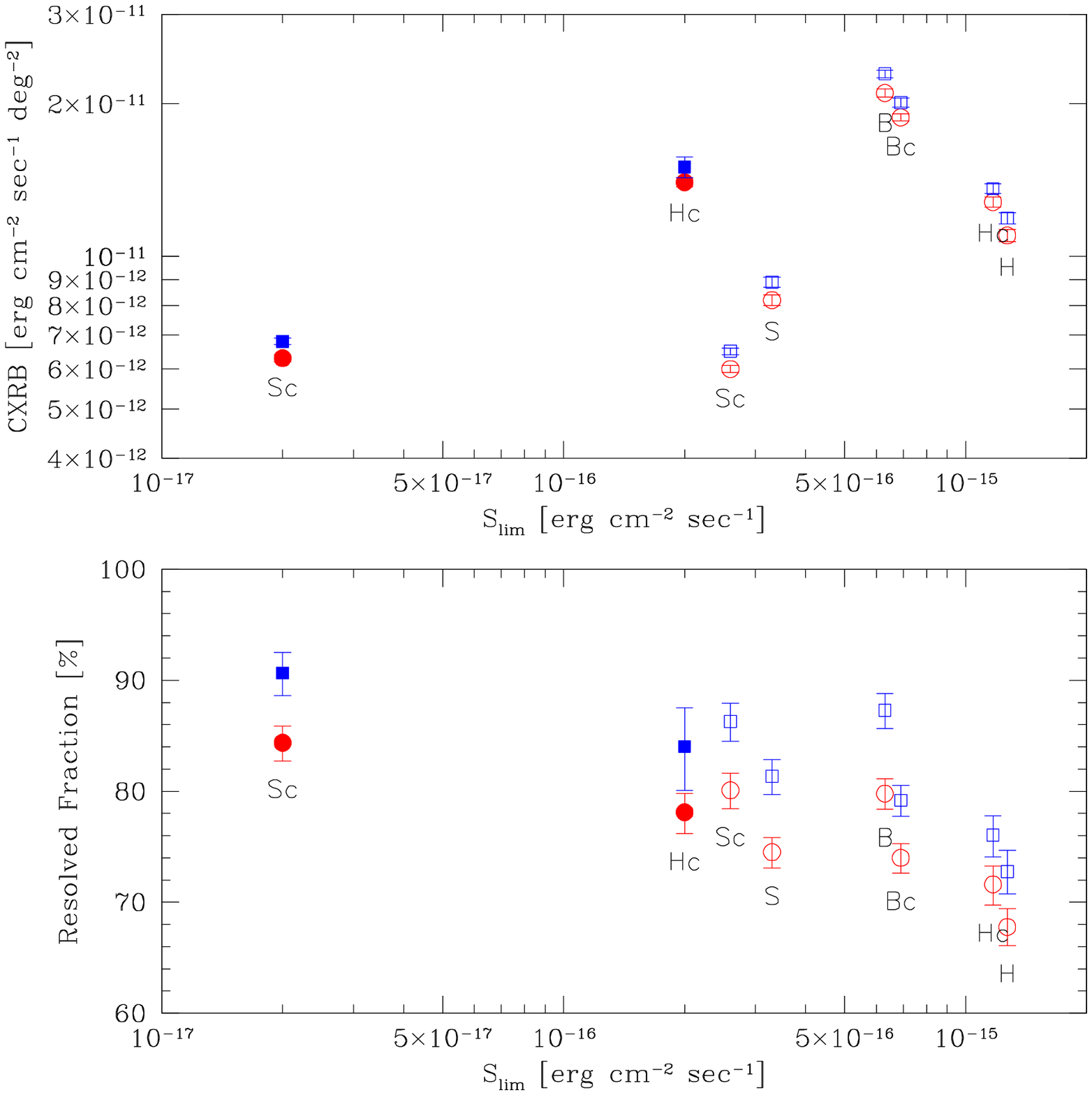}
\vspace{-10mm}
\figcaption[f21.eps]{
$top.$ The resolved CXRB flux density from the ChaMP ($open$ $symbols$) 
and the ChaMP+CDFs ($filled$ $symbols$)
number counts with target ($blue$ $squares$) and 
without target ($red$ $circles$) 
as a function of faint flux limit in 6 energy bands. 
$bottom.$ 
The percentage of the total CXRB flux density that is 
provided by the resolved sources of the ChaMP and ChaMP+CDFs
samples. 
The total CXRB in the Sc and Hc bands are from B04.
In other bands, the total CXRB was derived by summing (Bc) or rescaling
(B, S, and H) with those in the Sc and Hc bands (see \S 6.1).
Symbols are the same as the top panel. 
For the definition of the energy band, see Table \ref{tbl-energy-def}.
\label{fig-fmin_cxrb}}



\begin{thebibliography}{}
\bibitem[Baldi et al.(2002)]{bal02} Baldi, A., Molendi, S., Comastri, A., Fiore, F., 
Matt, G., \& Vignali, C. 2002, \apj, 564, 190  
\bibitem[Barkhouse et al.(2006)]{bar06}
Barkhouse, W. A., Green, P. J., Vikhlinin, A., Kim, D.-W., Perley, D.,
Cameron, R., Silverman, J., Mossman, A., Burenin, R., Jannuzi, B. T.,
Kim, M., Smith, M. G., Smith, R. C., Tananbaum, H., \& Wilkes, B. J.
2006, astro-ph/0603521, accepted to \apj
\bibitem[Basilakos et al.(2005)]{bas05} Basilakos, S., Plionis, M., 
Georgakakis, A., \& Georgantopoulos, I. 2005, MNRAS, 356, 183
\bibitem[Bauer et al.(2004)]{bau04} Bauer, F. E., Alexander, D. M.,
Brandt, W. N., Schneider, D. P., Treister, E., Hornschemeier, A. E., Garmire, G. P.
2004, \aj, 128, 2048 (B04) 
\bibitem[Brandt and Hasinger(2005)]{bra05} Brandt, W. N. \& Hasinger, G., 2005, 
\araa, 43, 827
\bibitem[Cappelluti et al.(2005)]{cap05} Cappelluti, N., Cappi, M., Dadina, M.,
Malaguti, G., Branchesi, M., D'Elia, V., \& Palumbo, G. G. C. 2005, A\&A, 430, 39
\bibitem[Cash(1979)]{cash79} Cash, W. 1979, \apj, 228, 939 
\bibitem[Cowie et al.(2002)]{cow02} Cowie, L. L., Garmire, G. P., Bautz, M. W., Barger, A. J., Brandt, W. N., \& Hornschemeier, A. E. 2002, \apj, 566, L5
\bibitem[Chiappetti et al.(2005)]{chi05} 
Chiappetti, L., Tajer, M., Trinchieri, G., Maccagni, D., Maraschi, L., 
Paioro, L., Pierre, M., Surdej, J., Garcet, O., Gosset, E., Le Fèvre, O., 
Bertin, E., McCracken, H. J., Mellier, Y., Foucaud, S., Radovich, M., 
Ripepi, V., Arnaboldi, M. 2005, A\&A, 439, 413
\bibitem[De Luca \& Molendi(2004)]{del04} De Luca, A., \& Moldendi, S. 2004, A\&A, 419, 837 
\bibitem[Gehrels et al.(1986)]{geh86} Gehrels, N., et al. 1986, \apj, 303, 336
\bibitem[Gendreau, Barcons, \& Fabian (1998)]{gen98} Gendreau, K. C., Barcons, X., 
\& Fabian, A. C. 1998, MNRAS, 297, 41
\bibitem[Green et al.(2004)]{gre04}
Green, P. J., Silverman, J. D., Cameron, R. A., Kim, D.-W., Wilkes, B. J.,
Barkhouse, W. A., LaCluyz, A., Morris, D., Mossman, A., Ghosh, H.,
Grimes, J. P., Jannuzi, B. T., Tananbaum, H., Aldcroft, T. L.,
Baldwin, J. A., Chaffee, F. H., Dey, A., Dosaj, A., Evans, N. R.,
Fan, X., Foltz, C., Gaetz, T., Hooper, E. J., Kashyap, V. L.,
Mathur, S., McGarry, M. B., Romero-Colmenero, E., Smith, M. G.,
Smith, P. S., Smith, R. C., Torres, G., Vikhlinin, A., Wik, D. R.
2004, \apjs, 150, 43
\bibitem[Harrison et al.(2003)]{har03} Harrison, F. A., Eckart, M. E., Mao, P. H., Helefand, D. J.,
\& Stern, D. 2003, \apj, 596, 944 (H03)
\bibitem[Hasinger et al.(1993)]{has93} 
Hasinger, G., Burg, R., Giacconi, R., Hartner, G., Schmidt, M., Trumper, J., Zamorani, G.
1993, A\&A, 275, 1  
\bibitem[Hickox \& Markevitch(2006)]{hic06} Hickox, R. C., \& Markevitch, M. 
2006, \apj, 645, 95 (HM06)
\bibitem[Johnson et al.(2003)]{joh03} Johnson, O., Best, P. N., \& Almaini, O. 
2003, MNRAS, 343, 924
\bibitem[Kenter and Murray(2003)]{ken03} Kenter, A. \& Murray, S. 2003, \apj, 584, 1016 
\bibitem[Kim, D.-W. and Fabbiano(2003)]{kim03} Kim, D.-W., \& 
Fabbiano, G. 2003, \apj, 586, 826
\bibitem[Kim, D.-W. et al.(2004a)]{kim04a} 
Kim, D.-W., Cameron, R. A., Drake, J. J., Evans, N. R., Freeman, P., Gaetz, T. J., 
Ghosh, H., Green, P. J., Harnden, F. R., Jr., Karovska, M., Kashyap, V., 
Maksym, P. W., Ratzlaff, P. W., Schlegel, E. M., Silverman, J. D., 
Tananbaum, H. D., Vikhlinin, A. A., Wilkes, B. J., \& Grimes, J. P.
2004a, \apjs, 150, 19 
\bibitem[Kim, D.-W. et al.(2004b)]{kim04b} 
Kim, D.-W., Wilkes, B. J., Green, P. J., Cameron, R. A., Drake, J. J., Evans, N. R., 
Freeman, P., Gaetz, T. J., Ghosh, H., Harnden, F. R., Jr., Karovska, M., 
Kashyap, V., Maksym, P. W., Ratzlaff, P. W., Schlegel, E. M., Silverman, J. D., 
Tananbaum, H. D., \& Vikhlinin, A. A. 2004b, \apj, 600, 59 (KD04)
\bibitem[Kim, E. et al.(2006)]{kim06b} Kim, E. et al. 2006, in preparation
\bibitem[Kim, M. et al.(2006)]{kim06a} Kim, M. et al. 2006, submitted to ApJS (KM06)
\bibitem[Kushino et al.(2002)]{kus02} 
Kushino, A., Ishisaki, Y., Morita, U., Yamasaki, N. Y., Ishida, M., Ohashi, T., Ueda, Y. 
2002, Publ. Astron. Soc. Japan, 54, 327 
\bibitem[La Franca et al.(2005)]{laf05} 
La Franca, F., Fiore, F., Comastri, A., Perola, G. C., Sacchi, N., Brusa, M., 
Cocchia, F., Feruglio, C., Matt, G., Vignali, C., Carangelo, N., Ciliegi, P., 
Lamastra, A., Maiolino, R., Mignoli, M., Molendi, S., Puccetti, S. 2005, \apj, 635, 864
\bibitem[Lumb et al.(2002)]{lum02} Lumb, D. H., Warwick, R. S., Page, M., \& De Luca, A. 2002, A\&A, 389, 93
\bibitem[Manners et al.(2003)]{man03}
Manners, J. C., Johnson, O., Almaini, O., Willott, C. J., Gonzalez-Solares, E., 
Lawrence, A., Mann, R. G., Perez-Fournon, I., Dunlop, J. S., McMahon, R. G., 
Oliver, S. J., Rowan-Robinson, M., Serjeant, S. 2003, MNRAS, 343, 293
\bibitem[Markevitch et al.(2003)]{mar03} 
Markevitch, M., Bautz, M. W., Biller, B., Butt, Y., Edgar, R., Gaetz, T., Garmire, G., 
Grant, C. E., Green, P., Juda, M., Plucinsky, P. P., Schwartz, D., Smith, R., Vikhlinin, A., 
Virani, S., Wargelin, B. J., Wolk, S. 2003, \apj, 583,70 
\bibitem[Moretti et al.(2003)]{mor03} Moretti, A., Campana, S., Lazzati, D., 
\& Tagliaferri, G. 2003, \apj, 588, 696 (M03)
\bibitem[Morrison \& McCammon (1983)]{mor83} Morrison, R.; McCammon, D. 1983, \apj, 270, 119
\bibitem[Murdoch et al.(1973)]{mur73} Murdoch, H. S., Crawford, D. F., 
\& Janucey, D. L. 1973, \apj, 183, 1
\bibitem[Mushotzky et al.(2000)]{mus00} Mushotzky, R. F., Cowie, L. L., Barger, A. J., \& Arnaud, K. A. 2000, Nature, 404, 459
\bibitem[Park et al.(2006)]{par06} 
Park, T., Kashyap, V. L., Siemiginowska, A., van Dyk, D. A., Zezas, A.,
Heinke, C., Wargelin, B. J. 2006, astro-ph/0606247, accepted to \apj
\bibitem[Parmer et al.(1999)]{par99} Parmer, A. N., Guainazzi, M., Oosterbroek T., Orr, A., Favata, F., 
Lumb, D., \& Malizia, A. 1999, A\&A, 345, 611
\bibitem[Press et al.(1992)]{pre92} Press, W. H., Teukolsky, S. A.,
Vetterling, W. T., Flannery, B. P 1992, \it{Numerial Recipes in Fortran},
\rm{Cambridge University Press}
\bibitem[Rosati et al.(2002)]{ros02} 
Rosati, P., Tozzi, P., Giacconi, R., Gilli, R., Hasinger, G., Kewley, L., Mainieri, V., 
Nonino, M., Norman, C., Szokoly, G., Wang, J. X., Zirm, A., Bergeron, J., Borgani, S., 
Gilmozzi, R., Grogin, N., Koekemoer, A., Schreier, E., Zheng, W.
2002, \apj, 566, 667 
\bibitem[Rosati et al.(1998)]{ros98} Rosati, P., Della Ceca, R., Norman, C., \&
Giacconi, R. 1998, \apj, 492, 21 
\bibitem[Silverman et al.(2005)]{sil05}
Silverman, J., Green, P., Barkhouse, W., Kim, D.-W., Aldcroft, T., Cameron, R.,
Wilkes, B., Mossman, A., Ghosh, H., Tananbaum, H., Smith, M.,
Smith, R., Smith, P., Foltz, C., Wik, D., Jannuzi, B. 2005, \apj, 618, 123
\bibitem[Stark et al.(1992)]{sta92} 
Stark, A. A., Gammie, C. F., Wilson, R. W., Bally, J., Linke, R. A., Heiles, C., Hurwitz, M.
1992, \apjs, 79, 77
\bibitem[Tozzi et al.(2006)]{toz06}
Tozzi, P., Gilli, R., Mainieri, V., Norman, C., Risaliti, G., Rosati, P.,
Bergeron, J., Borgani, S., Giacconi, R., Hasinger, G., Nonino, M., Streblyanska, A.,
Szokoly, G., Wang, J. X., \& Zheng, W. 2006, A\&A, 451, 457
\bibitem[Vikhlinin et al.(1995)]{vik95} Vikhlinin, A., Forman, W., Jones, C., \& Murray, S.
1995, \apj, 451, 553 
\bibitem[Worsley et al.(2005)]{wor05}
Worsley, M. A., Fabian, A. C., Bauer, F. E., Alexander, D. M., Hasinger, G.,
Mateos, S., Brunner, H., Brandt, W. N., \& Schneider, D. P., 2005, \mnras, 357, 1281
\bibitem[Yang et al.(2004)]{yan04} Yang, Y., Mushotzky, R. F., Steffan, A. T., Barger, A. J., \& Cowie, L. L.
2004, \aj, 128, 1501
\end{thebibliography}
\end{document}